\documentclass{aastex631}

\shorttitle{Hierarchical Structure and Type 4 Scaling Relations}
\shortauthors{He et al.}

\graphicspath{{./}{figures/}}
\begin{document}

\title{Hierarchical Dynamics within Molecular Clouds: Type 4 Scaling Relations of Molecular Clouds in the Second Galactic Quadrant}

\correspondingauthor{Xuepeng Chen, Yuehui Ma}
\email{xpchen@pmo.ac.cn, mayh@pmo.ac.cn}

\author[0009-0004-2947-4020]{Suziye He}
\affiliation{Purple Mountain Observatory and Key Laboratory of Radio Astronomy, Chinese Academy of Sciences, 10 Yuanhua Road, Nanjing 210033, People's Republic of China\\}
\affiliation{School of Astronomy and Space Science, University of Science and Technology of China, 96 Jinzhai Road, Hefei 230026, People's Republic of China\\}

\author[0000-0002-8051-5228]{Yuehui Ma}
\affiliation{Purple Mountain Observatory and Key Laboratory of Radio Astronomy, Chinese Academy of Sciences, 10 Yuanhua Road, Nanjing 210033, People's Republic of China\\}

\author[0000-0003-3151-8964]{Xuepeng Chen}
\affiliation{Purple Mountain Observatory and Key Laboratory of Radio Astronomy, Chinese Academy of Sciences, 10 Yuanhua Road, Nanjing 210033, People's Republic of China\\}
\affiliation{School of Astronomy and Space Science, University of Science and Technology of China, 96 Jinzhai Road, Hefei 230026, People's Republic of China\\}

\author[0000-0003-0746-7968]{Hongchi Wang}
\affiliation{Purple Mountain Observatory and Key Laboratory of Radio Astronomy, Chinese Academy of Sciences, 10 Yuanhua Road, Nanjing 210033, People's Republic of China\\}
\affiliation{School of Astronomy and Space Science, University of Science and Technology of China, 96 Jinzhai Road, Hefei 230026, People's Republic of China\\}

\author[0000-0002-6388-649X]{Miaomiao Zhang}
\affiliation{Purple Mountain Observatory and Key Laboratory of Radio Astronomy, Chinese Academy of Sciences, 10 Yuanhua Road, Nanjing 210033, People's Republic of China\\}

\author[0009-0009-3431-1150]{Zhenyi Yue}
\affiliation{Purple Mountain Observatory and Key Laboratory of Radio Astronomy, Chinese Academy of Sciences, 10 Yuanhua Road, Nanjing 210033, People's Republic of China\\}
\affiliation{School of Astronomy and Space Science, University of Science and Technology of China, 96 Jinzhai Road, Hefei 230026, People's Republic of China\\}

\author[0009-0009-8707-8620]{Xiangyu Ou}
\affiliation{Purple Mountain Observatory and Key Laboratory of Radio Astronomy, Chinese Academy of Sciences, 10 Yuanhua Road, Nanjing 210033, People's Republic of China\\}
\affiliation{School of Astronomy and Space Science, University of Science and Technology of China, 96 Jinzhai Road, Hefei 230026, People's Republic of China\\}

\begin{abstract}
We present a statistical study of the hierarchical structure of molecular clouds in the second quadrant of the Galactic midplane ($139\fdg75 \le l \le 159\fdg75$, $-5\fdg25 \le b \le 8\fdg25$), based on high-sensitivity $^{12}$CO and $^{13}$CO $J=1-0$ data from the Milky Way Imaging Scroll Painting survey. Using DBSCAN and a non-binary Dendrogram algorithm, we identify 2,912 $^{13}$CO clouds. Type 2 (multicloud, single-tracer) scaling relations show arm-dependent trends: the power-law index of the linewidth–size relation is $\sim 0.32$ in the Local Arm and $\sim 0.42$ in the Perseus and Outer Arms. The classical Keto–Heyer relation is weak in the Local Arm but present in the Perseus and Outer Arms with a slope of $\sim 0.3$. The virial parameter is anti-correlated with cloud size, mass, and surface density, consistent with previous surveys. Type 4 (single-cloud, single-tracer) scaling relations exhibit stronger internal correlations. The linewidth–size index spans $\sim0.3$–$0.8$ and correlates positively with sonic Mach number, indicating that the relation is not universal but depends on cloud compressibility. While substructures do not follow the conventional surface-density-based Keto–Heyer relation, a volume-density formulation restores a clear virial scaling, with evolutionary tracks aligning along lines of constant virial parameter. Moreover, the structural complexity of molecular clouds, as characterized by the relationship between the number of substructures and the min$\_$delta parameter, correlates with surface density rather than sonic Mach number, contrary to results from turbulence-only magnetohydrodynamic simulations. Together, these results are more consistent with a gravity-regulated hierarchical dynamical picture than with a purely turbulence-supported scenario.

\end{abstract}



\section{Introduction} \label{sec:intro}

Molecular clouds have been observed to exhibit supersonic velocity dispersion for decades \citep{Wilson1970}, yet the physical origin of the supersonic linewidth remains a topic of debate \citep{Vazquez2025}. Since molecular clouds are the cradles of star formation, at least a portion (if not all) of the gas within molecular clouds is in a state of gravitational collapse. Therefore, a natural hypothesis is that the supersonic linewidth originates from the global gravitational collapse of the molecular cloud \citep{Goldreich1974}. However, this hypothesis quickly encounters a problem: if the entire molecular cloud collapsed within a free-fall timescale and formed stars, the observed star formation rate (SFR) of the Milky Way would be two orders of magnitude higher than what is currently observed \citep{Zuckerman1974a, Zuckerman1974b}. For decades, the prevailing view—hereafter the turbulent support (TS) model—proposed that the linewidth is a manifestation of supersonic turbulence within molecular clouds providing support against self-gravity (e.g., \citealp{VS1999, MacLow2004, BP2006, McKee2007, Hennebelle2012}). Only on the scale of cores can self-gravity potentially dominate in this picture. Therefore, only a small fraction of the mass of a molecular cloud undergoes gravitational contraction, resulting in a low SFR. On the other hand, the Global Hierarchical Collapse (GHC) model proposed by \cite{Vazquez2019} suggests that molecular clouds remain gravitationally dominated on all scales. However, the gravitational collapse within them is not a singular, monolithic, and synchronous process, but rather a hierarchical and multi-scale phenomenon. Turbulence within molecular clouds is only transonic and insufficient to counteract gravitational collapse in these structures. These two scenarios imply fundamentally different dynamical states of molecular clouds, which may be reflected in differences in the statistical relations among their observable properties.

Several empirical scaling relations have been identified among the physical properties of molecular clouds. For example, \cite{Larson1981} obtained a power-law relationship between the physical scale and the velocity dispersion of molecular clouds, that is, $\sigma_v \propto R^{\eta}$, with $\eta=0.38$, while later surveys have suggested values around 0.5 (e.g., \citealp{Solomon1987, Rice2016, Miville2017, Hacar2023}). The linewidth-size relation is regarded as a result of the energy cascade of turbulence within the so-called inertial sub-range \citep{McKee2007}, while the power-law index $\eta=0.5$ aligns well with the prediction of \textit{Burgers'} turbulence. 
However, the linewidth of a molecular cloud may depend not only on its size but also on its surface density \citep{Keto1986}.
\cite{Heyer2009} showed that giant molecular clouds tend to lie along the locus corresponding to a virial parameter of unity in the Keto–Heyer diagram, yielding the relation $\sigma_v/R^{0.5} \propto \Sigma^{0.5}$.
Within the TS framework, this behavior is interpreted as evidence that molecular clouds are approximately in virial equilibrium, whereas the GHC model attributes it to approximate energy conservation during near–free-fall collapse \citep{BP2011}.
These contrasting interpretations imply different scaling behaviors for molecular clouds and their substructures, motivating large-sample statistical tests.

According to the tracers used and the sample characteristics, \citet{Goodman1998} categorized the linewidth-size relations into four types, each representing different physical interpretations. For example, the original work of \cite{Larson1981} made use of observations of multiple tracers toward numerous molecular clouds, which is classified as type 1 (multitracer, multicloud intercomparison). The work of \cite{Heyer2009} is regarded as type 2, since it exclusively utilized $^{13}$CO data to determine the physical properties of multiple molecular clouds. Studies based on molecular cloud samples, i.e., type 1 and type 2 \citep{Solomon1987, Rice2016, Miville2017}, fail to illustrate whether such linewidth-size relations still hold within individual molecular clouds. Meanwhile, type 3 studies employ various tracers with different critical densities to identify hierarchical structures within a single cloud. For example, \cite{Traficante2020} utilized the emission of the $\rm N_2H^+(1-0)$ line to investigate the kinematic states of a set of clumps while using $\rm ^{12}CO \ and\ ^{13}CO$ to determine their parental filaments; \cite{Wu2010} used tracers including $\rm CS(7-6), \ CS(2-1), \ HCN(1-0),\ and \ C^{34}S(5-4)$ to derive velocity dispersion toward many cores and examine a ``modified type 3 relation''.
Intercomparison across multiple tracers may introduce systematic effects, including selection biases and mismatches between tracers. As pointed out by \cite{Goodman1998}, the so-called type 4 scaling relation, i.e., a single-tracer study of a single cloud, provides the most direct probe of the internal dynamics within a given density regime. Therefore, a large-sample statistical analysis of the type 4 linewidth–size relation can effectively assess whether the velocity dispersions of substructures within molecular clouds are consistent with the predictions of \textit{Burgers'} turbulence.

Due to the requirement of using a single spectral line to identify hierarchical structures within a cloud, studies of type 4 linewidth–size relations remain relatively scarce. In early work, such as \citet{Goodman1998}, different intensity thresholds within a single cloud were adopted to identify structures at various hierarchical levels. \citet{Rosolowsky2008} later introduced the Dendrogram method to identify structures on different scales and investigated the type 4 linewidth–size relation in the L1448 star-forming region. Meanwhile, approaches that do not rely on explicit structure identification have also been developed. The use of Principal Component Analysis (PCA) to investigate the linewidth–size relation \citep{Heyer2004} is regarded by some researchers as a form of type 4 analysis \citep{Yun2021}. In the PCA approach, the linewidth–size relation is derived on a pixel-by-pixel basis, which may be affected by multiple velocity components along the line of sight. Therefore, we adopt the Dendrogram method to identify hierarchical structures.

The original Dendrogram algorithm is a binary-branching scheme, in which each branch splits into two substructures, often producing numerous “phantom branches” \citep{Storm2014}. Based on this framework, we have modified the algorithm to construct a non-binary Dendrogram and have applied it to both the Maddalena cloud \citep{Shen2024} and the Rosette cloud \citep{He2026}. To date, nearly all type 4 studies have been limited to case analyses, and a systematic statistical investigation of type 4 scaling relations based on an optically thin tracer in large samples is still lacking.

In this work, we utilize the non-binary Dendrogram algorithm and the $^{13}$CO $J=1-0$ isotopic spectral lines from $l=139^{\circ}.75$ to $l=159^{\circ}.75$ and $b=-5^{\circ}.25$ to $b=8^{\circ}.25$ of the Milky Way Imaging Scroll Painting (MWISP) project to investigate several type 4 scaling relations among molecular clouds. The $^{13}$CO $J=1-0$ emission line serves as an excellent tracer for the type 4 scaling relation. As a moderately optically thin transition from a CO isotopologue--CO being the second most abundant molecule in the interstellar medium--this line probes both the denser regions within molecular clouds and their more diffuse envelopes. The selection of this region is motivated by several considerations. Existing $^{13}$CO surveys in this area remain limited, and the molecular clouds have not yet been systematically identified or characterized. Furthermore, as part of the outer Galaxy, this region is less affected by velocity crowding, allowing more reliable structure identification in Position–Position–Velocity (PPV) space using the non-binary Dendrogram algorithm and therefore enabling a robust investigation of the type 4 scaling relations.

The paper is organized as follows. In Section \ref{sec:2}, we give a brief introduction to the observational data and the structural identification algorithm. In Section \ref{sec:3.1}, we study the distributions and correlations of the physical properties of molecular clouds using a type 2 approach. In Section \ref{sec:3.2}, we focus on the type 4 scaling relations of substructures within single molecular clouds. We discuss the relations between type 2 and type 4 scaling relations and propose a revised Keto-Heyer relation in Section \ref{sec:4}, and finally present a summary of our results in Section \ref{sec:5}.

\section{Data and Method} \label{sec:2}
\subsection{Observation and Data Reduction} \label{sec:2.1}
We have observed the $^{12}$CO, $^{13}$CO, and C$^{18}$O $J=1-0$ emission toward the Galactic plane, with a sky coverage of $20^\circ \times 13.5^\circ$, from $l=139^{\circ}.75$ to $l=159^{\circ}.75$ and $b=-5^{\circ}.25$ to $b=8^{\circ}.25$. These observations are conducted with the Purple Mountain Observatory (PMO) 13.7 m millimeter-wavelength telescope in Delingha, China, and is a part of the MWISP project \citep{Yang2026}, which is an unbiased survey of the $J=1-0$ line emission of the three isotopologues of carbon monoxide toward the northern Galactic plane. The data used in this work within Galactic latitudes of $|b|\leq5^{\circ}.25$ are from the MWISP Phase I survey, observed between 2013 and 2022, and are publicly available \citep{Yang2026}. Data within Galactic latitudes of $5^{\circ}.25<b<8^{\circ}.25$ were observed from 2022 to 2025 as part of the ongoing MWISP Phase II survey.

The PMO-13.7 m telescope is equipped with a $3 \times 3$ multibeam sideband-separating superconducting spectroscopic array receiver (SSAR) system \citep{Shan2012}, allowing simultaneous observation of $^{12}$CO, $^{13}$CO, and C$^{18}$O $J=1-0$ emissions. The half-power beamwidth (HPBW) of the telescope is approximately $52^{\prime \prime}$ at 110 GHz and $50^{\prime \prime}$ at 115 GHz, respectively, while the data are regridded to a pixel size of $30^{\prime\prime} \times 30^{\prime\prime}$. The backend of the SSAR comprises 18 Fast Fourier Transform Spectrometer (FFTS) containing 16,384 channels with a bandwidth of 1 GHz, which provides a velocity resolution of $\rm 0.16\ km\ s^{-1}$ at the frequency of 115 GHz of the $^{12}$CO $J=1-0$ line and $\rm 0.17 \ km \ s^{-1}$ at 110 GHz of the $\rm ^{13}CO $ and C$^{18}$O $J=1-0$ lines. 

The raw data are preprocessed following the preliminary analysis of noise characteristics in \citet{Cai2021}, which includes flagging bad channels, mitigating edge effects, correcting baseline distortions, and removing line contamination. The final RMS noise level, $\sigma_\mathrm{RMS}$, for the $^{12}$CO data is $\rm \sim 0.5\ K$ per $\rm 0.16\ km\ s^{-1}$ and $\rm \sim 0.3\ K $ per $\rm 0.17\ km\ s^{-1}$ for the $^{13}$CO and C$^{18}$O data, respectively. 

\subsection{Structure Identification} \label{sec:2.2}
We employ a non-binary Dendrogram algorithm, which is a modified version of the \texttt{astrodendro} \citep{Robitaille2019} implementation of the hierarchical Dendrogram algorithm of \cite{Rosolowsky2008}, to identify $^{13}$CO clouds and their hierarchical substructures. The Dendrogram algorithm is a widely used structure identification algorithm that can represent the hierarchical features in an n-dimensional data cube as a tree diagram.
However, the original Dendrogram algorithm is rarely applied directly for hierarchical structure identification in practice, which is most likely due to its binary-branching nature, i.e., a branch structure always splits into two substructures, thereby generating physically insignificant ``phantom branches'' \citep{Storm2014}. In our previous work, we developed a non-binary Dendrogram algorithm based on the original Dendrogram and applied it to the Maddalena molecular cloud and Rosette molecular cloud separately (\citealp{Shen2024, He2026}). Details on modification to the algorithm and influence on scaling relations are presented in \citet{He2026}.

In outer galaxy studies, we typically define a single-connected PPV volume in the $^{12}$CO emission data as a $^{12}$CO cloud. The $^{13}$CO structures contained within such a $^{12}$CO cloud are referred to as the associated $^{13}$CO clouds. The identification of $^{12}$CO clouds is accomplished using an approach developed by \cite{Yan2020}, which is based on the DBSCAN algorithm \citep{Ester1996}. \citet{Yan2020} systematically examined the effects of MinPts, and emission thresholds on DBSCAN cloud identification using MWISP data and showed that the resulting cloud population and statistical properties are generally robust against reasonable parameter variations. We therefore adopt their recommended settings ($\epsilon =1$, MinPts=4, and T$_{\rm cutoff}=2\sigma_{\rm RMS})$ for the identification of parent $^{12}$CO clouds. As suggested by \cite{Yan2020}, several post-selection criteria are employed to avoid contamination caused by noise and bad channels: (1) number of voxels $\geq 16$; (2) peak main-beam brightness temperature $\geq 5\sigma_\mathrm{RMS}$; (3) presence of more than one compact $2 \times 2$ pixel region ($\rm 1^\prime \times 1^\prime$) in the projection; and (4) the number of velocity channels $\geq 3$. Molecular clouds identified by DBSCAN which fail to meet the above criteria will be removed from the raw catalog. Through the process mentioned above, a total of 10,071 $^{12}$CO clouds are derived from the region. 

For each $^{12}$CO molecular cloud, we get its corresponding $^{13}$CO data cube and then apply the non-binary Dendrogram algorithm to each $^{13}$CO data cube with the following parameters: (1) min$\_$value$=2\sigma_\mathrm{RMS}$, voxels with brightness temperature less than min$\_$value will be excluded from the identification; (2) min$\_$delta$=3\sigma_\mathrm{RMS}$, the minimum length threshold for leaf structures to be considered independent and able to merge with other structure; (3) min$\_$npix$=27$, the minimum volume threshold (in voxels) for leaf structures to be considered independent, this constraint also applies to branch structures in the non-binary edition; (4) branch$\_$delta$=\sigma_\mathrm{RMS}$, the minimum length threshold for branch structures to be considered independent, which is newly added in the non-binary Dendrogram.

\begin{figure*}[!htb]
	\centering
	\begin{minipage}[t]{0.49\linewidth}
		\centering
		\includegraphics[width=\linewidth]{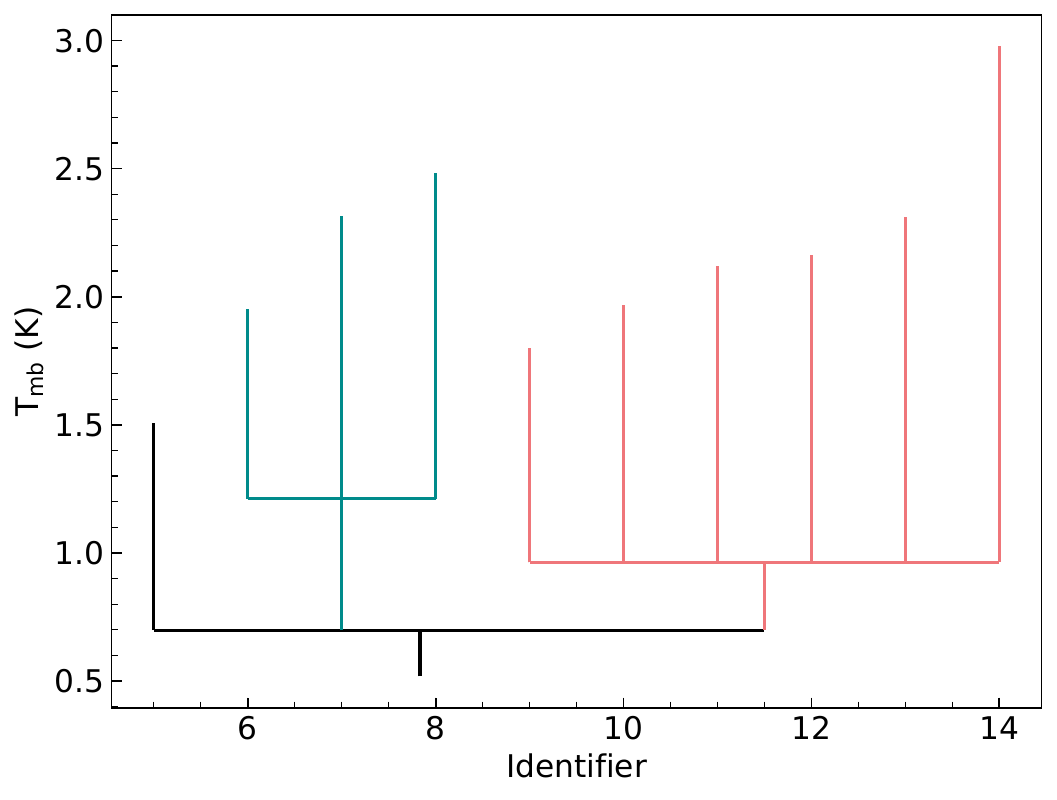}
		\\(a)
	\end{minipage}
	\begin{minipage}[t]{0.49\linewidth}
		\centering
		\includegraphics[width=\linewidth]{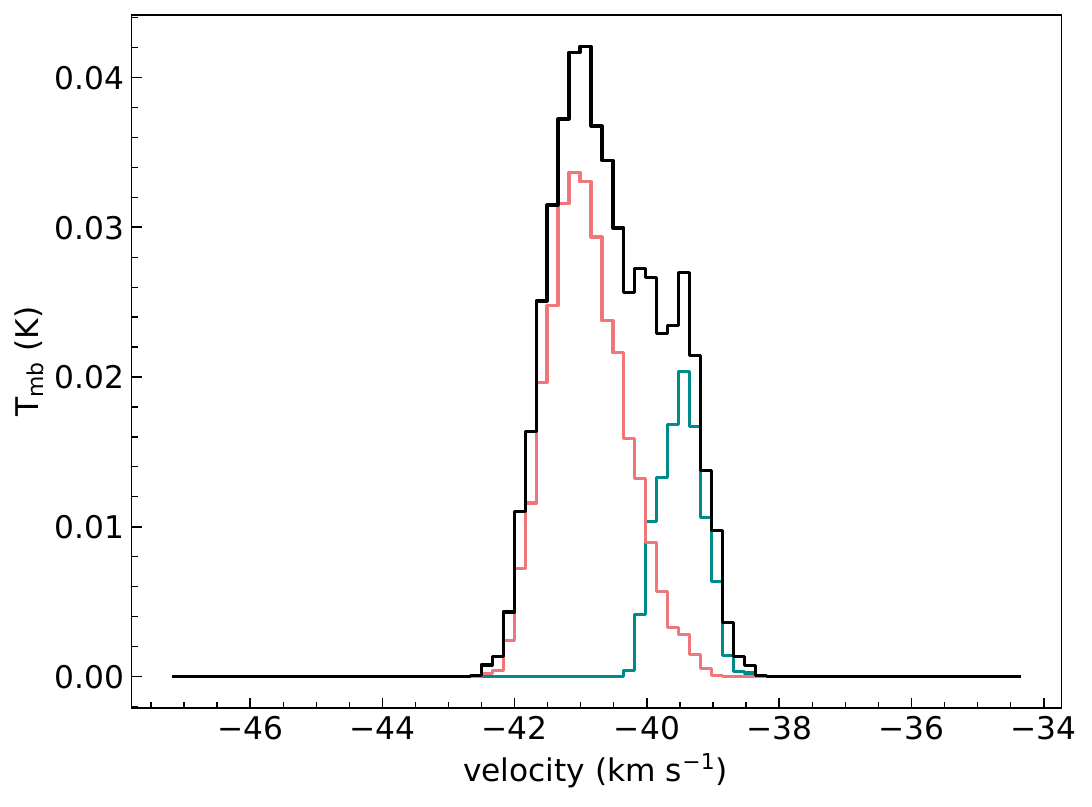}
		\\(b)
	\end{minipage}
	\caption{Schematic diagram of the decomposition of a Dendrogram trunk into single-velocity-component structures. (a) The tree diagram of the original trunk (black) and its substructures, the decomposed single-component branches are colored in red and green, respectively. (b) Average spectra of the structures with corresponding color in panel (a).}\label{fig:decompose}
\end{figure*}

The structures located at the lowest level of the hierarchical tree diagram, as identified by the Dendrogram algorithm, are referred to as ``trunks''. As the Dendrogram algorithm identifies structures solely based on connectivity in PPV space, a trunk structure may comprise multiple gaussian components, which can bias the estimation of its velocity dispersion. To address this issue, we further examine the $^{13}$CO average spectrum of each trunk structure. If the average $^{13}$CO spectrum of a trunk exhibits multiple velocity components, we separate the trunk into distinct velocity components and treat each as an individual $^{13}$CO cloud. A schematic diagram is presented in Figure \ref{fig:decompose}. The left panel shows a tree diagram of a trunk structure (black), which further splits into two single-velocity-component structures (red and green). The corresponding average spectra are shown in the right panel using the same colors. We hereafter refer to these single-velocity-component structures as $^{13}$CO clouds, which are defined as kinematic substructures nested within the parent $^{12}$CO clouds. After excluding all molecular clouds that intersect with the field boundary, a total of 2,912 $^{13}$CO clouds within 10,071 $^{12}$CO clouds are identified through the above process. In the following, all statistical analyses are performed using these 2,912 $^{13}$CO clouds unless otherwise specified. The physical parameters of the 2,912 $^{13}$CO clouds are presented in Table \ref{tab:example}.

\begin{deluxetable}{lcccccccccccc}
  \tablecaption{$^{13}$CO Cloud Catalog\label{tab:example}}
  \tablehead{
    \colhead{Index} & \colhead{$l$} & \colhead{b} & \colhead{$\sigma_\mathrm{maj}$} & \colhead{$\sigma_\mathrm{min}$} & \colhead{PA$^{a}$} & \colhead{$v$} & \colhead{$\sigma_v$} & \colhead{$d$} & \colhead{$R$} & \colhead{$M$} & \colhead{$\alpha_{vir}$} & \colhead{arm} \\
    \colhead{} & \colhead{(deg)} & \colhead{(deg)} & \colhead{(arcsec)} & \colhead{(arcsec)} & \colhead{(deg)} & \colhead{(m s$^{-1}$)} & \colhead{(m s$^{-1}$)} & \colhead{(kpc)} & \colhead{(pc)} & \colhead{(M$_\odot$)} & \colhead{} & \colhead{} \\
  }
  \startdata
    1 & 147.567 & 1.271 & 57.276 & 35.274 & 168.816 & -13077.446 & 257.121 & 1.19 & 0.494 & 5.135 & 7.401 & loc \\
    2 & 148.480 & 0.024 & 94.681 & 26.732 & 82.571 & -43687.540 & 280.447 & 4.81 & 2.240 & 237.160 & 0.864 & per \\
    3 & 148.523 & 0.131 & 43.268 & 31.478 & 130.154 & -43222.010 & 313.262 & 4.74 & 1.620 & 112.028 & 1.650 & per \\
    4 & 139.945 & -0.574 & 109.856 & 50.044 & 159.825 & -39746.323 & 333.438 & 3.45 & 2.368 & 326.446 & 0.938 & per \\
    5 & 139.795 & -0.662 & 44.337 & 35.777 & -167.899 & -39017.093 & 302.887 & 3.36 & 1.241 & 105.472 & 1.255 & per \\
    6 & 140.145 & -0.562 & 45.615 & 29.795 & 171.773 & -36985.555 & 336.416 & 3.17 & 1.084 & 26.616 & 5.356 & per \\
    7 & 148.420 & -3.619 & 63.598 & 44.123 & 68.564 & -24179.634 & 332.024 & 2.29 & 1.125 & 39.602 & 3.640 & per \\
    \multicolumn{13}{c}{$\vdots$} \\
    2910 & 143.800 & -0.001 & 65.529 & 27.822 & -149.033 & -14587.717 & 377.297 & 1.25 & 0.493 & 5.848 & 13.954 & loc \\
    2911 & 143.787 & -0.056 & 59.975 & 31.877 & 173.472 & -14499.000 & 356.145 & 1.24 & 0.502 & 4.863 & 15.212 & loc \\
    2912 & 149.268 & 6.778 & 29.800 & 25.654 & 138.622 & -10630.512 & 149.431 & 1.02 & 0.261 & 1.547 & 4.383 & loc \\
  \enddata
  \tablecomments{Table \ref{tab:example} is published in its entirety in the machine-readable format.
      A portion is shown here for guidance regarding its form and content.\\
      $^a$ The position angle is the angle of the major axis measured counter-clockwise from the positive x-axis.}
\end{deluxetable}

\subsection{Physical Properties of Molecular Clouds}\label{subsec:2.3}
\subsubsection{Distance Estimation}\label{subsubsec:2.3.1}
We adopt kinematic distances for individual molecular clouds. Since our analysis mainly focuses on type 4 scaling relations within individual molecular clouds, the absolute values of the cloud distances have little impact on our results, as long as the internal structures are located at approximately the same distance. 

 We employ two methods to estimate kinematic distances. The first is the \texttt{kd} Python package \citep{Wenger2018}, which uses a traditional rotation-curve-based approach with the updated solar motion parameters from \citet{Reid2019}. The second is the parallax-based Bayesian distance estimator of \citet{Reid2019}, which incorporates spiral-arm information.

For sources assigned a kinematic distance below 200 pc by the \texttt{kd} Python package, we adopt the distance estimate from the Bayesian method. While it is important to note that this combined approach still carries significant uncertainties, employing other, more accurate distance-determination techniques—such as extinction-based methods—is impractical for our unbiased molecular cloud sample, as they cannot be uniformly applied to all objects in the catalog. Distance uncertainties may contribute scatter to the type 2 scaling relations presented in Section \ref{subsec:3.1.3}, because the physical cloud properties depend on the adopted distances. In contrast, for the type 4 scaling relations analyzed in Section \ref{subsec:3.2.1}—the primary focus of this study—the absolute distance uncertainty does not affect the derived power-law indices but only shifts the normalization of the relations.

\subsubsection{Physical Parameters}
For identified structures, the Dendrogram algorithm provides a \texttt{ppv\_catalog} function to derive their physical properties. Specifically, the angular radius of each structure is given by
\begin{equation}
    R_a = \eta \sqrt{\sigma_{maj}\sigma_{min}},
\end{equation}
where $\sigma_{maj}$ and $\sigma_{min}$ represent the rms size along the major and minor axes of the fitted ellipse for the 2D projection of the structure on the sky plane, respectively, and $\eta$ is the conversion factor from the one-dimensional rms size to the three-dimensional sphere radius. We adopt the recommended value of $\eta = 1.91$ from \cite{Rosolowsky2008}. The physical radius is then given by $R = R_a \times D$, where $D$ is the distance of the cloud. Beam deconvolution is not applied to the structure radii in this work, as a simple Gaussian deconvolution is inappropriate for the structures at higher levels of the dendrogram, which are defined by high brightness temperature contours. A statistical analysis of the cloud sample shows that, for 99.6\% of the structures, the change in radius after deconvolution is less than 20\%, and for 88.7\%, the change is less than 10\%. The impact of beam deconvolution on the derived cloud properties and scaling relations is therefore negligible.

The column density of a structure is calculated under the assumption that the $^{13}$CO molecules are in local thermodynamic equilibrium (LTE) \citep{Dickman1978, Mangum2015}. Then the column density of $^{13}$CO is given by 
\begin{equation}
    N(^{13}\mathrm{CO}) = 2.42\times10^{14}\frac{1+0.88/T_\mathrm{ex}}{1-exp(-5.29/T_\mathrm{ex})}T_\mathrm{ex}\int \tau(^{13}\mathrm{CO})d\nu,
\end{equation}
where $T_\mathrm{ex}$ is the excitation temperature of $^{13}$CO, and $\tau(^{13}\mathrm{CO})$ is the optical depth of the $^{13}$CO $J=1-0$ emission. Assuming the $^{12}$CO $J=1-0$ emission is optically thick, we derive the excitation temperature of $^{12}$CO from the optically thick $^{12}$CO emission, and further assume that $^{12}$CO and $^{13}$CO share the same excitation temperature. The column density of $\rm H_{2}$ can be obtained using the abundance ratio of $\rm [H_2/^{12}CO]=1.1\times 10^4$ \citep{Frerking1982} and $\rm [^{12}CO/^{13}CO]=6.21 \times {d_\mathrm{GC}}+18.71$ \citep{Milam2005}, where $d_\mathrm{GC}$ is the galactocentric distance of the molecular cloud, in units of kpc. It can be derived from the galactic coordinates of the cloud and the galactocentric distance of the Sun, for which we adopt $R_\odot = 8.15$ kpc \citep{Reid2019}. The mass of a molecular cloud can be derived by integrating its column density:
\begin{equation}
    M = \mu_\mathrm{H_2} m_\mathrm{H} D^2 \int N(\mathrm{H_2}) d\Omega,
\end{equation}
where $\mu_\mathrm{H_2} = 2.8$ is the mean molecular weight per hydrogen molecule, $m_\mathrm{H}$ is the mass of a hydrogen atom, and $\Omega$ represents the solid angle of the cloud. 

The velocity dispersion, $\sigma_v$, of each structure is defined as the intensity-weighted second moment of the velocity over all voxels within the structure. Under the LTE assumption, we use the excitation temperature as an approximation of the kinetic temperature, $T_\mathrm{kin}$. The non-thermal velocity dispersion can be derived with $\sigma_\mathrm{nt}=\sqrt{\sigma_v^2-\sigma_\mathrm{th}^2}$, where $\sigma_\mathrm{th}$ is the one-dimensional thermal velocity dispersion of the $^{13}$CO molecules. The sonic Mach number $M_\mathrm{s}$ of a cloud is defined as $M_\mathrm{s}=\sigma_{\rm nt}/c_\mathrm{s}$, where $c_\mathrm{s}=\sqrt{k_\mathrm{B} T_{\rm kin}/\mu_\mathrm{p} m_{\rm H}}$ is the isothermal sound speed, and $\mu_\mathrm{p}=2.37$ is the mean molecular weight per free particle \citep{Kauffmann2008}.

A commonly used indicator of the gravitational binding state of a cloud is its virial parameter \citep{Bertoldi1992}, which is defined as

\begin{equation}
    \alpha_\mathrm{vir} \equiv \frac{2 E_\mathrm{k}}{|E_\mathrm{g}|} \approx \frac{5\sigma_\mathrm{nt}^2R}{GM},
\end{equation}

where $G$ is the gravitational constant. For a spherically symmetric molecular cloud in gravitational equilibrium, i.e., a Bonnor-Ebert sphere \citep{Bonnor1956}, the mass threshold for maintaining hydrostatic equilibrium is
\begin{equation}
    M_\mathrm{BE} = 2.43 \frac{\sigma_v^2 R}{G},
\end{equation}
resulting in a critical virial parameter of around 2. Therefore, we empirically consider molecular clouds with a virial parameter less than 2 to be gravitationally bound.

We would like to clarify that in this work, We adopt the bijection scheme rather than the extrapolation paradigm for the Dendrogram structures \citep{Rosolowsky2006, Rosolowsky2008}. In the bijection scheme, the properties of a structure are measured directly from the emission assigned to it, whereas the extrapolation paradigm estimates the properties that would be obtained by extending the structure boundary from $T_\mathrm{edge}$ to $T_\mathrm{edge} = 0$ K. \citet{Rosolowsky2008} showed that the bijection scheme is generally more appropriate for the analysis of hierarchical substructures within molecular clouds. Since the Type 4 scaling relations in this work are constructed using structures at all hierarchical levels, including leaves, branches, and trunks, we adopt the bijection scheme uniformly for all structures to ensure consistency in the derived physical properties and scaling relations.

\section{Results} \label{sec:results}
\begin{figure*}[!htb]
	\centering
	\includegraphics[width=\textwidth]{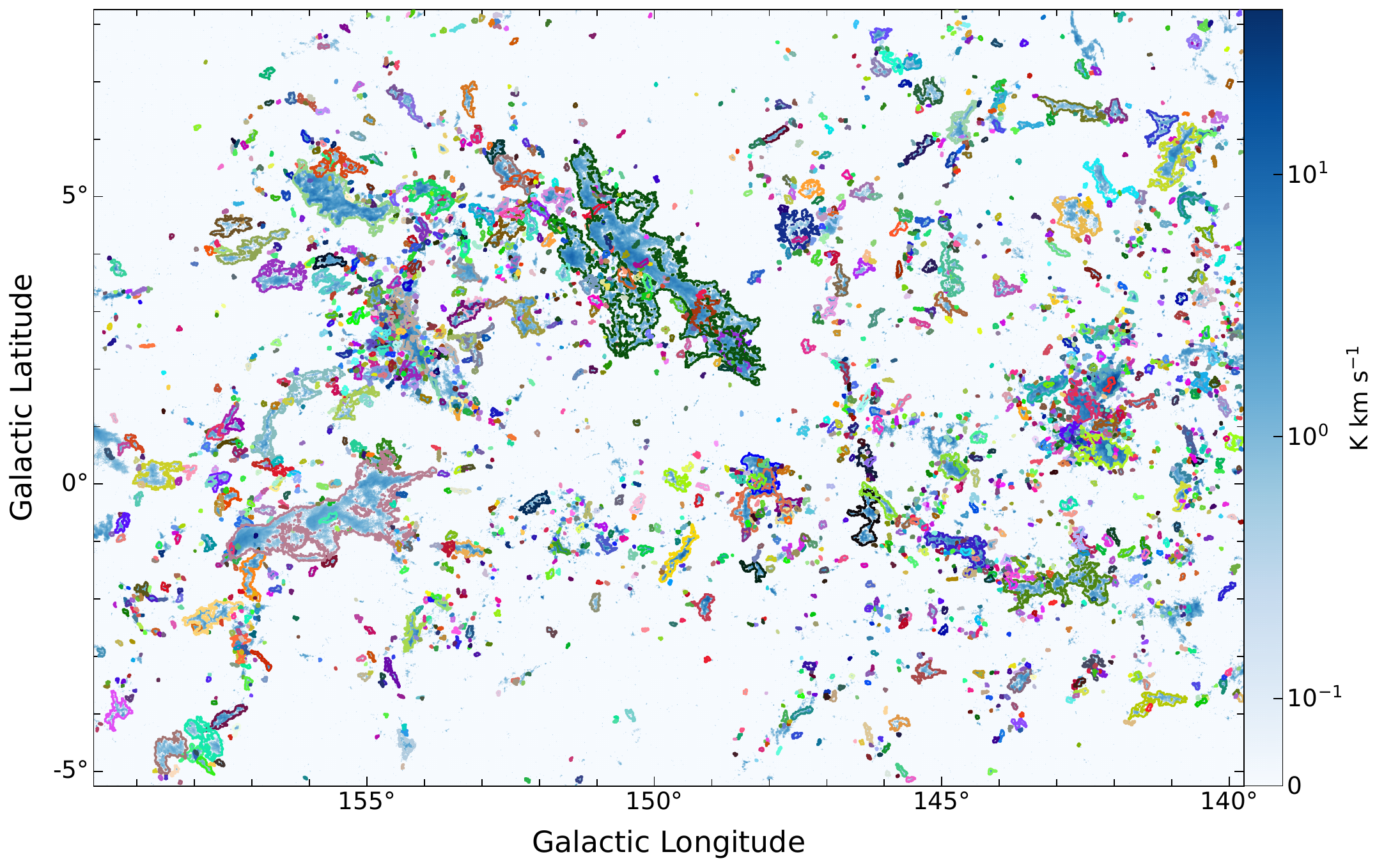}
	\caption{Integrated intensity maps of $\rm ^{13}$CO emission, in units of $\rm K\ km\ s^{-1}$. The contours present the projected boundaries of the identified $^{13}$CO clouds.}\label{fig:rgb}
\end{figure*}

\begin{figure*}[!htb]
	\centering
	\includegraphics[width=\textwidth]{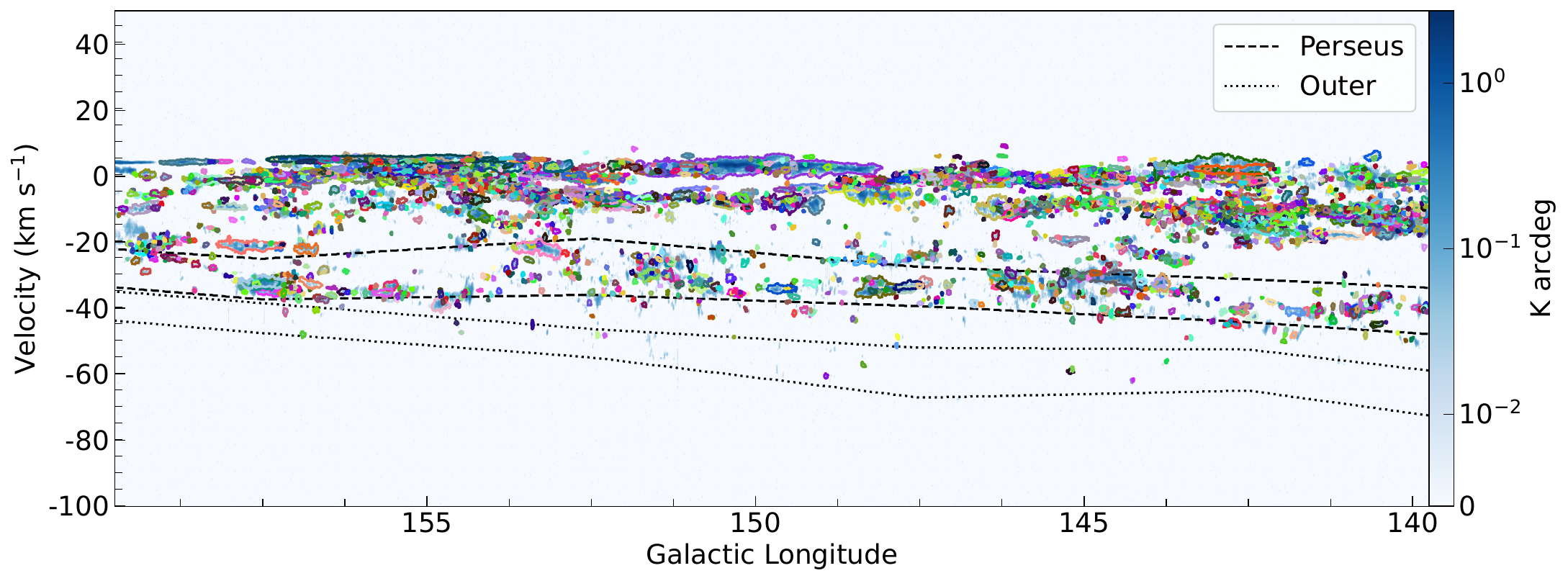}
	\caption{$l-v$ diagram of $^{13}$CO emission. The black dashed line and dotted line present the velocity range at the specific Galactic longitude of the Perseus and Outer Arms given by \cite{Sun2024}.}\label{fig:lv}
\end{figure*}

\subsection{Multicloud Analysis} \label{sec:3.1}
\subsubsection{Overall Distribution of $^{13}$CO Clouds} \label{sec:3.1.1}
Figure \ref{fig:rgb} shows the integrated intensity map of $^{13}$CO $J=1-0$ emission, where contours of different colors mark the boundaries of identified $^{13}$CO clouds. The position-velocity distribution of the $^{13}$CO emission along the direction of Galactic longitude is presented in Figure \ref{fig:lv}. The black dashed and dotted lines present the projected velocity ranges corresponding to different galactic longitudes of the Perseus and Outer Arms given by \cite{Sun2024}. As shown in Figure \ref{fig:lv}, the observed $^{13}$CO emission is predominantly concentrated near the Local and Perseus Arms, with only a small portion extending to the Outer Arm. Within most of the observed region of this study, the Local Arm can be kinematically separated into two layers: the upper layer in the $l-v$ diagram is the Gould Belt Layer, while the lower layer is the Cam OB1 Layer \citep{Du2017}.

Using the method illustrated in Section \ref{sec:2.1}, a total of 2,912 $^{13}$CO clouds have been identified. Each cloud is then assigned to one spiral arm according to its centroid velocity and the $l-v$ ranges of the spiral arms. Specifically, we first used the velocity ranges of the Perseus Arm and Outer Arm at each Galactic longitude obtained by \citet{Sun2024} (i.e., the dashed and dotted lines in Figure \ref{fig:lv}) to assign clouds whose centroid velocities fall within these velocity ranges to the corresponding arm. For clouds whose centroid velocities do not fall within the adopted velocity ranges of the Perseus or Outer Arms, we assign them to the spiral arm whose central velocity is closest to the cloud centroid velocity. The central velocity of the Local Arm is adopted from \citet{Reid2019}. Among all the 2,912 molecular clouds, 2,210 are assigned to the Local Arm, 673 to the Perseus Arm, and 29 to the Outer Arm.

\subsubsection{Physical Properties of Identified $^{13}$CO Clouds}

The Dendrogram algorithm can directly provide various properties of molecular clouds, including centroid position, angular radius, flux, centroid velocity, velocity dispersion. By integrating these properties with the kinematic distances obtained through the method described in Section \ref{subsubsec:2.3.1}, we can determine the effective radius of the molecular clouds. Furthermore, utilizing the $^{12}$CO and $^{13}$CO emissions under the LTE assumption, we can derive the mass of each molecular cloud, which allows us to calculate additional parameters, such as the virial parameter. Details of the derivation of the physical properties are provided in Section \ref{subsec:2.3}.

\begin{figure*}[!htb]
	\centering
	\begin{minipage}[t]{0.49\linewidth}
		\centering
		\includegraphics[width=\linewidth]{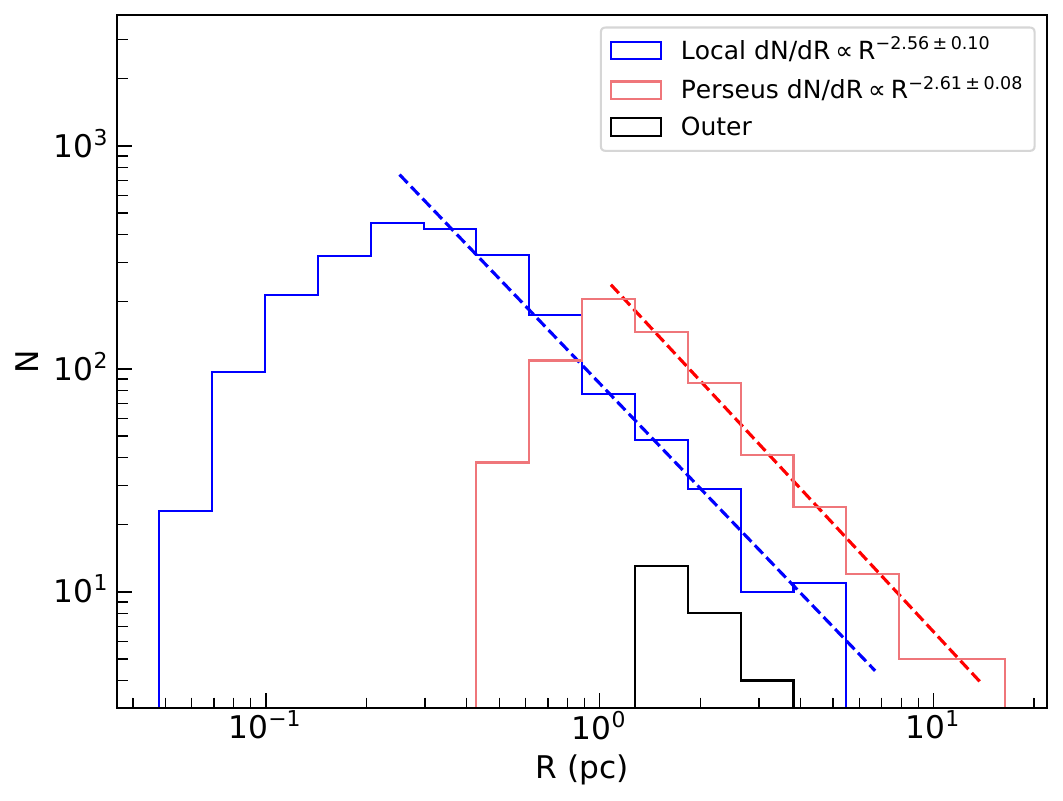}
		\\(a)
	\end{minipage}
	\begin{minipage}[t]{0.49\linewidth}
		\centering
		\includegraphics[width=\linewidth]{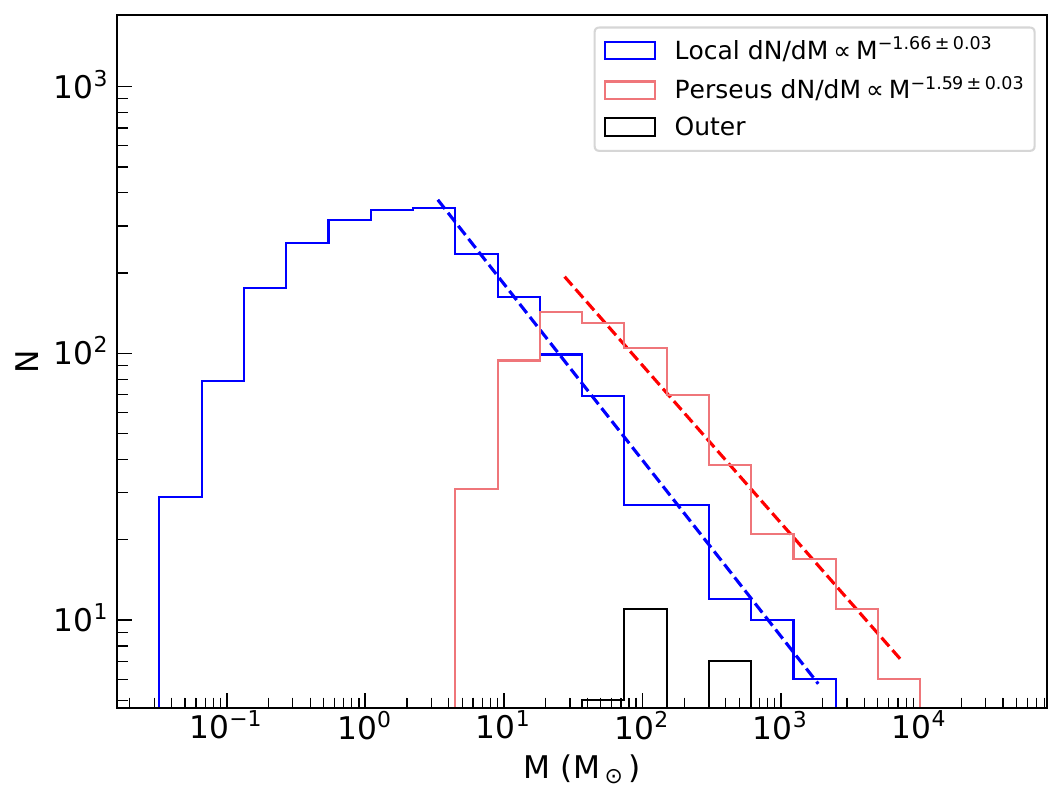}
		\\(b)
	\end{minipage}
	\centering
	\begin{minipage}[t]{0.49\linewidth}
		\centering
		\includegraphics[width=\linewidth]{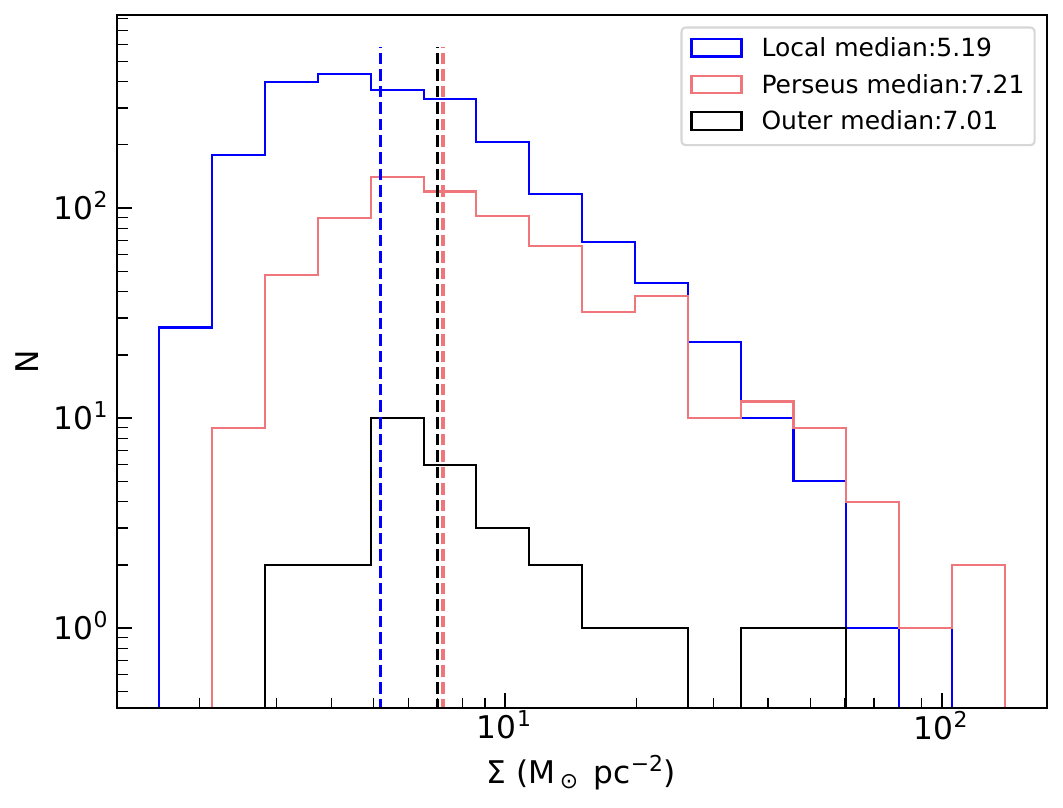}
		\\(c)
	\end{minipage}
	\begin{minipage}[t]{0.49\linewidth}
		\centering
		\includegraphics[width=\linewidth]{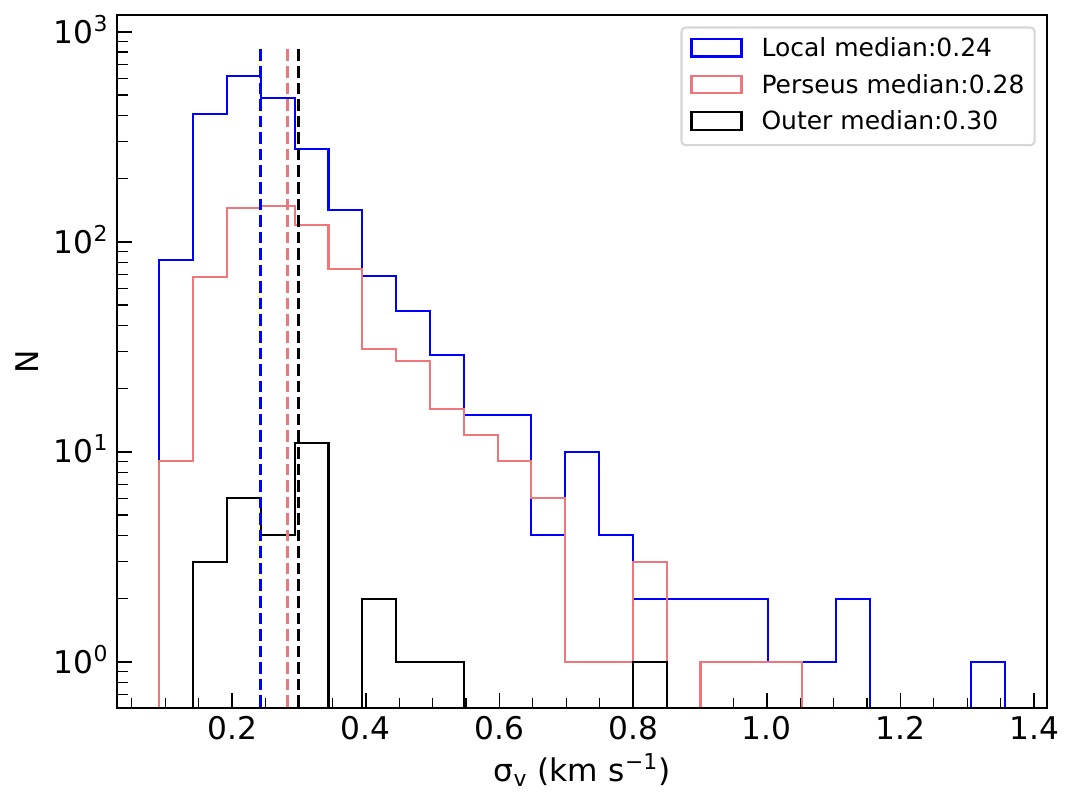}
		\\(d)
	\end{minipage}
	\caption{Histograms of (a) radius, (b) mass, (c) surface density, and (d) velocity dispersion for the identified $^{13}$CO clouds. Columns colored in blue, red, and black represent molecular clouds within Local, Perseus, and Outer Arms, respectively. Dashed lines show the power-law fits in panels (a) and (b); while present the median surface density and velosity dispersion in panels (c) and (d).}
	\label{fig:histograms}
\end{figure*}

Figure \ref{fig:histograms} shows the histograms of the radius, mass, surface density, and velocity dispersion of the identified $^{13}$CO clouds, with different colors representing molecular clouds located in different spiral arms.
The distributions of the radius and mass of the $^{13}$CO clouds are presented in panels (a) and (b), respectively. For each distribution, we fit a power-law relation for the high value end, and the fitting results are presented with lines in corresponding colors. Due to the limited number of molecular clouds in the Outer Arm, fitting results for structures in this arm are not provided. Our fittings give the relations of $dN/dR \propto R^{-2.56 \pm 0.10}$ and $dN/dM \propto M^{-1.66 \pm 0.03}$ in the Local Arm and $dN/dR \propto R^{-2.61 \pm 0.08}$ and $dN/dM \propto M^{-1.59 \pm 0.03}$ in the Perseus Arm. The power-law indices of the mass and radius functions for Local Arm and Perseus Arm are generally consistent. Our results differs from those of previous studies. For instance, \citet{Ma2021} utilized the MWISP data in the second Galactic quadrant and obtained $dN/dR \propto R^{-2.54 \pm 0.25}$ for $^{13}$CO clouds in the Perseus Arm and $dN/dR \propto R^{-1.78 \pm 0.17}$ in the Local Arm. Separately, \citet{Roman2010} employed $^{13}$CO emission from the Galactic Ring Survey (GRS) in the first quadrant, finding $dN/dR \propto R^{-4.9 \pm 0.65}$ for molecular clouds with radii greater than 10 pc, and $dN/dM \propto M^{-2.64 \pm 0.25}$ for molecular clouds with masses exceeding $10^5 M_\odot$.

The histogram of the surface density of molecular clouds is shown in Figure \ref{fig:histograms}(c), and the dashed lines indicate the corresponding median surface density in each spiral arm, which are 5.19, 7.21, and 7.01 $\rm M_\odot \ pc^{-2}$ for molecular clouds in the Local, Perseus, and Outer Arms, respectively. Despite having comparable surface density distributions across the Local, Perseus, and Outer Arms, molecular clouds in the Local Arm show a modestly lower median surface density. Accordingly, the minimum surface density in each arm also increase with galactocentric distance, with values of 1.62, 2.33, and 3.32 $\rm M_\odot \ pc^{-2}$ for the Local, Perseus, and Outer Arms, respectively. Using data from the MWISP survey in the second Galactic quadrant, \citet{Ma2021} found that the surface densities of $^{13}$CO clouds all exceed 10 $\mathrm{M}_\odot \ \mathrm{pc}^{-2}$, which are systematically higher than the molecular clouds in our sample. The surface density of molecular clouds in our sample is much lower than the median surface density of 144 $\rm M_\odot \ pc^{-2}$ obtained for molecular clouds in the first quadrant by the GRS, and also lower than the median of 73 $\rm M_\odot \ pc^{-2}$ obtained for the inner Galaxy molecular cloud sample from the SEDIGISM survey \citep{Duarte2021}. This phenomenon is to be expected. As noted by \citet{Heyer2015}, the molecular hydrogen surface density in the inner Galaxy is inherently much higher than that in the outer Galaxy. Furthermore, these results are also influenced by differences in the tracers used (e.g., $^{13}$CO $J=2-1$ employed by the SEDIGISM survey) and the cloud identification algorithms applied (e.g., CLUMPFIND used by the GRS.)

The histogram of the velocity dispersion of molecular clouds is shown in Figure \ref{fig:histograms}(d). The distributions are narrow, with dynamical ranges of $\rm\sim 0.09 \ to \ 1.36 \ km \ s^{-1}$ in the Local Arm, $\rm \sim 0.11 \ to \ 1.04 \ km \ s^{-1}$ in the Perseus Arm, and $\rm\sim 0.18 \ to \ 0.84 \ km \ s^{-1}$ in the Outer Arm. The median velocity dispersions of the $\rm^{13}CO$ molecular clouds in this work are $\sim0.24$, $\sim0.28$, and $\rm\sim0.30\ km \ s^{-1}$ for the Local, Perseus, and Outer Arms, respectively, which are significantly smaller than those given by the CfA $\rm ^{12}CO $ $J=1-0$ survey (e.g., $\rm \sim 2.8\ km \ s^{-1}$ from \citealp{Miville2017} and $\rm\sim 2.1\ km \ s^{-1}$ from \citealp{Rice2016}). This discrepancy likely arises because the molecular clouds in our sample have smaller spatial extents than those in theirs due to a different structure identification procedure, and also because the $\rm ^{12}CO$ emission may be affected by optical depth broadening. The median velocity dispersions of our molecular clouds align well with previous results derived with DBSCAN using MWISP $^{13}$CO data \citep{Dong2023}. 

\begin{figure*}[!htb] 
	\centering
	\includegraphics[width=0.7\textwidth]{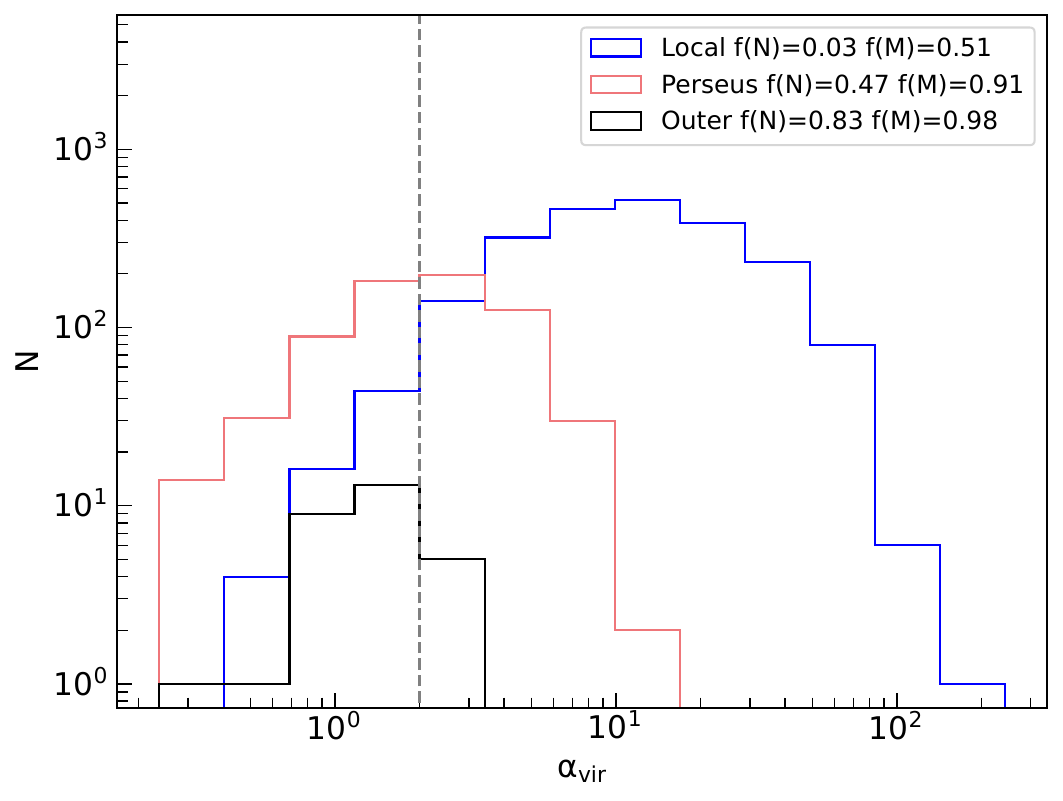}
	\caption{Distribution of the $\alpha_\mathrm{vir}$ of the identified $^{13}$CO clouds with different colors indicating three spiral arms. The vertical grey dashed line stands for $\alpha_\mathrm{vir} = 2$, and structures with smaller virial parameter are considered as gravitationally bound. The number fraction and mass fraction of self-gravitationally bound structures are labeled in the legend as \( f(N) \) and \( f(M) \), respectively.}\label{fig:hist_vp}
\end{figure*}

Figure \ref{fig:hist_vp} presents the distribution of the $\alpha_\mathrm{vir}$ for the $^{13}$CO clouds. Different colors indicate different spiral arms, and structures with virial parameters less than 2 are considered gravitationally bound. 
Among the 2,210 molecular clouds identified in the Local Arm, only 64 (3\%) are gravitationally bound. Despite their low abundance, these bound molecular clouds account for 51\% of the total mass in the region. A similar trend is observed in the Perseus Arm, where 317 out of 673 molecular clouds (47\%) are bound, contributing 91\% of the total mass. In the Outer Arm, 24 out of 29 molecular clouds (83\%) are gravitationally bound, representing 98\% of the mass.

When considering all 2,912 molecular clouds across these regions, only 14\% are gravitationally bound, yet they dominate the mass budget, contributing 82\% of the total. This mass fraction is significantly higher than the value of 19\% reported by \cite{Miville2017} using $^{12}$CO data.

It is worth noting that the differences between molecular clouds in different spiral arms should be interpreted in the context of observational selection effects. Because the survey has a nearly uniform sensitivity and angular resolution, the same observational limits correspond to different physical scales and mass sensitivities at different distances. As a result, diffuse, low-mass, and/or small-linewidth clouds are more difficult to identify in the more distant spiral-arm samples than in the Local Arm sample. Therefore, the differences in physical properties and scaling relations among different spiral-arm subsamples represent statistical trends measured within the observational selection limits.

\subsubsection{Scaling Relations} \label{subsec:3.1.3}
In this section, we present the type 2 scaling relations—specifically, the linewidth-size relation, mass-size relation, Keto-Heyer relation, and the relation between $\sigma_v$ and $\Sigma R$—of the obtained $^{13}$CO clouds. 

\begin{figure*}[!htb]
	\centering
	\begin{minipage}[t]{0.49\linewidth}
		\centering
		\includegraphics[width=\linewidth]{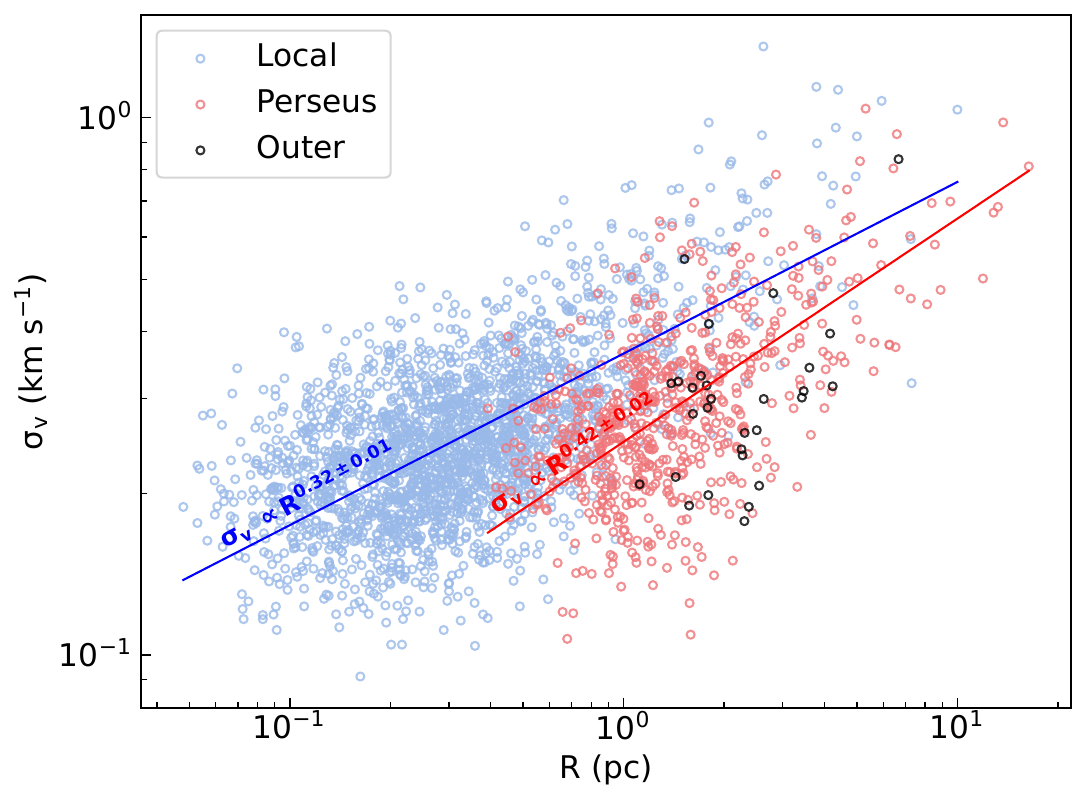}
		\\(a)
	\end{minipage}
	\begin{minipage}[t]{0.49\linewidth}
		\centering
		\includegraphics[width=\linewidth]{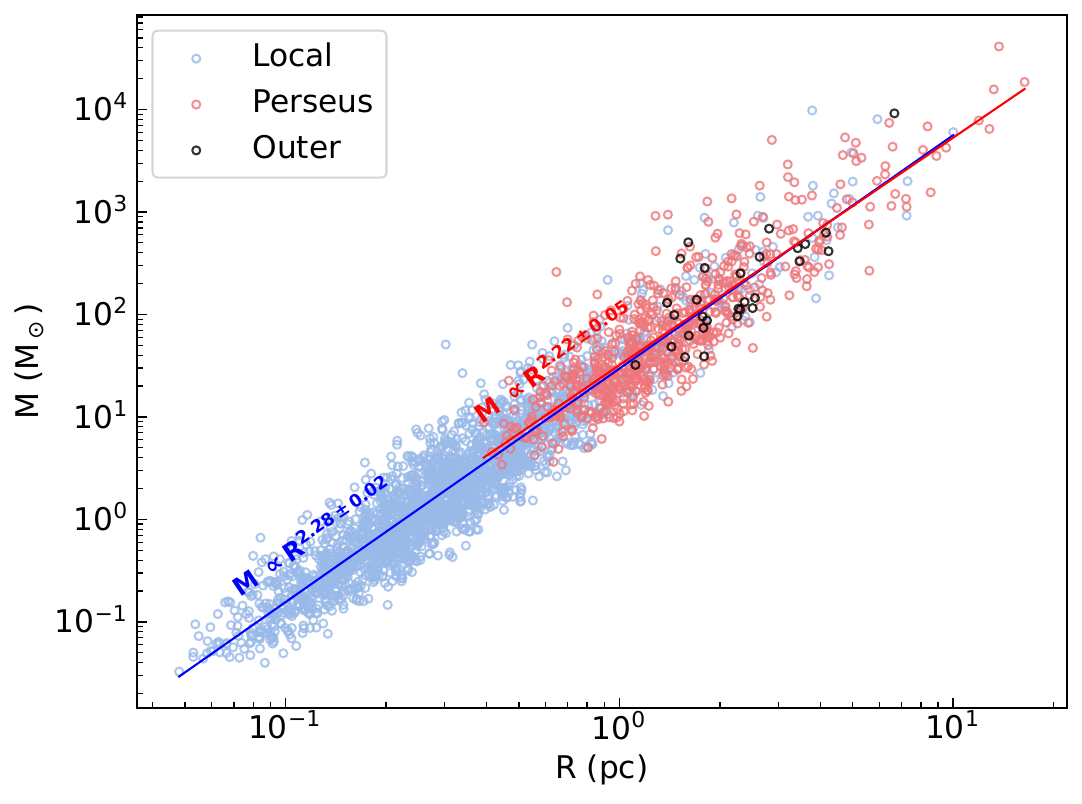}
		\\(b)
	\end{minipage}
	\begin{minipage}[t]{0.49\linewidth}
		\centering
		\includegraphics[width=\linewidth]{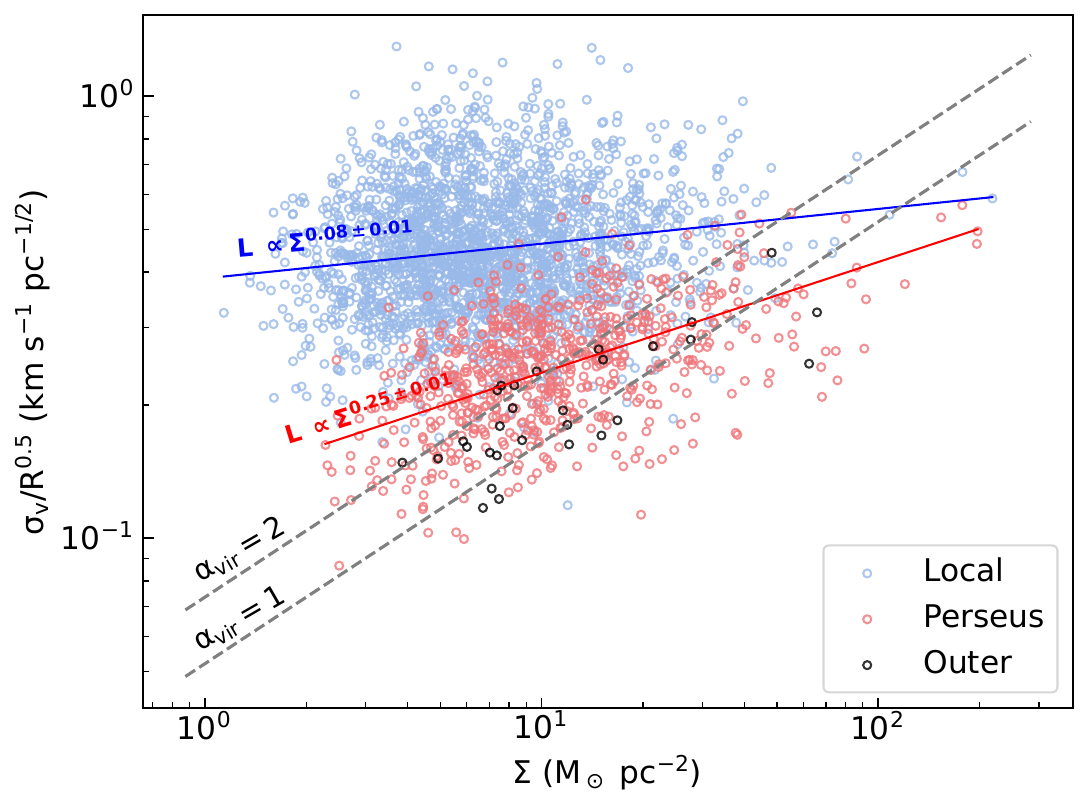}
		\\(c)
	\end{minipage}
	\begin{minipage}[t]{0.49\linewidth}
		\centering
		\includegraphics[width=\linewidth]{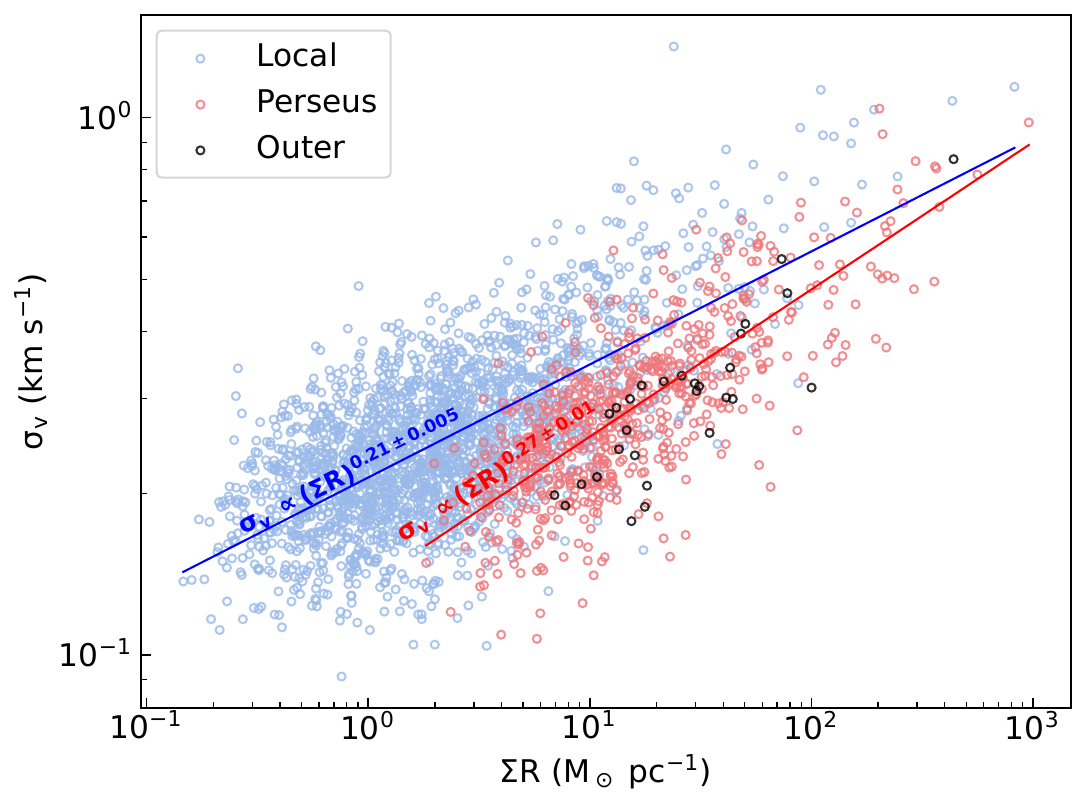}
		\\(d)
	\end{minipage}
	\caption{(a) Linewidth-size relation, (b) mass-size relation, (c) Keto-Heyer relation, and (d) $\sigma_v-\Sigma R$ relation of $^{13}$CO clouds. Structures assigned to the Local, Perseus, and Outer arm are indicated with blue, red, and black, respectively. The blue solid line indicates our best fit for molecular clouds in the Local Arm, and structures in the Perseus and Outer Arms are fitted as a whole (red solid line) as they show little difference in their distribution. }\label{fig:scaling relation}
\end{figure*}

The linewidth–size relation of the identified $^{13}$CO molecular clouds in our region is presented in Figure \ref{fig:scaling relation}(a). Molecular clouds in the Perseus and Outer Arms show overlapping distributions in mass and size (Figure \ref{fig:histograms}(a)(b)), and no clear separation in the $\sigma_v$–$R$ plane. We therefore combine these two arms into a single sample when fitting the scaling relations. The Local Arm clouds are fitted separately (blue line), while the combined Perseus and Outer Arm sample is shown in red.
Molecular clouds in the Perseus and Outer Arms show a relation of $\sigma_v \propto R^{0.42\pm 0.02}$, whereas the index drops to $0.32 \pm 0.01$ for molecular clouds in the Local Arm. 
Such a difference has also been reported in other statistical studies using MWISP $^{13}$CO data \citep{Ma2021}. This may indicate a varying relative importance of turbulence in molecular clouds at different Galactocentric distances. However, as discussed above, the comparison among different spiral-arm samples may also be affected by distance-dependent selection effects, since the same sensitivity and angular resolution correspond to different physical limits at different distances.

Figure \ref{fig:scaling relation}(b) presents the mass-size relation of the molecular clouds, giving the relation of $M\propto R^{2.28 \pm 0.02}$ for molecular clouds in the Local Arm and $M\propto R^{2.22 \pm 0.05}$ in the Perseus and Outer Arms. Among the four scaling relations, the $M-R$ relation exhibits the highest degree of consistency across different spiral arms, and the power-law indices we obtained align well with previous studies \citep{Roman2010, Miville2017, Ma2021, Dong2023}. 

Figure \ref{fig:scaling relation}(c) shows the Keto-Heyer relation, i.e., correlation between the Larson ratio, $\mathcal{L}\equiv \sigma_v/R^{0.5}$, and surface density, $\Sigma$, of molecular clouds. Molecular clouds in the Local Arm show only a weak correlation and yield $\mathcal{L} \propto \Sigma^{0.08 \pm 0.01}$ with a Pearson correlation coefficient of 0.12. For molecular clouds in the Perseus and Outer Arms, the Pearson coefficient increases to 0.51 and the power-law index increases to $\rm 0.25 \pm 0.01$, which is still shallower than the slope of 0.5 given by \cite{Heyer2009}. Molecular clouds in the Perseus and Outer Arms tend to have significantly smaller virial parameters compared to those in the Local Arm at similar column densities. Furthermore, structures in the Local Arm are clustered in regions far from the dashed lines representing virial parameters of 1 or 2. If the trend is not due to the relatively high uncertainty of kinematic distance for molecular clouds in the Local Arm, this may suggest that molecular clouds in the Local Arm require additional confinement, such as external pressure, to be in a state of virial equilibrium \citep{Bertoldi1992}. 

Figure \ref{fig:scaling relation}(d) illustrates the relation between $\sigma_v$ and $\Sigma R$ for molecular clouds. Among the three scaling relations involving $\sigma_v$, the $\sigma_v - \Sigma R$ relation shows the tightest correlation, with Pearson correlation coefficients of 0.66 for molecular clouds in the Local Arm and 0.76 for those in the Perseus and Outer Arms. Using a different tracer, \cite{Miville2017} reached the same conclusion with the linear fitting of $\sigma_v = 0.23 (\Sigma R)^{0.43\pm 0.14}$ for their Gaussian decomposition-derived $^{12}$CO clouds in the Galaxy. In our study, molecular clouds in the Local Arm give the relation of $\sigma_v \propto (\Sigma R)^{0.21\pm0.005}$ while those in the Perseus and Outer Arms yield $\sigma_v \propto (\Sigma R)^{0.27\pm0.01}$. The smaller power-law indices are consistent with the lower indices observed in the Keto-Heyer relation.

\subsubsection{Variation of Virial Parameter} \label{subsec:3.1.4}
Although molecular clouds are believed to be in near virial equilibrium, systematic variations in their virial parameter have been widely observed. Giant molecular clouds with the largest masses are often subvirial ($\alpha_\mathrm{vir} < 2$), while clumps with the lowest-column densities and/or masses tend to appear supervirial ($\alpha_\mathrm{vir} > 2$). \citet{Krumholz2019} proposed that this behavior naturally arises if molecular clouds and their substructures follow the linewidth–size relation. Assuming $\sigma_v \propto R^{0.5}$ and $M \propto R^3$ (i.e., constant volume density), it follows that $\alpha_{\rm vir} \propto R^{-1}$ and $\alpha_{\rm vir} \propto M^{-1/3}$. In this section, we examine the variation of the virial parameter with other properties of molecular clouds in our sample.

\begin{figure*}[!htb]
	\centering
	\begin{minipage}[t]{0.32\linewidth}
		\centering
		\includegraphics[width=\linewidth]{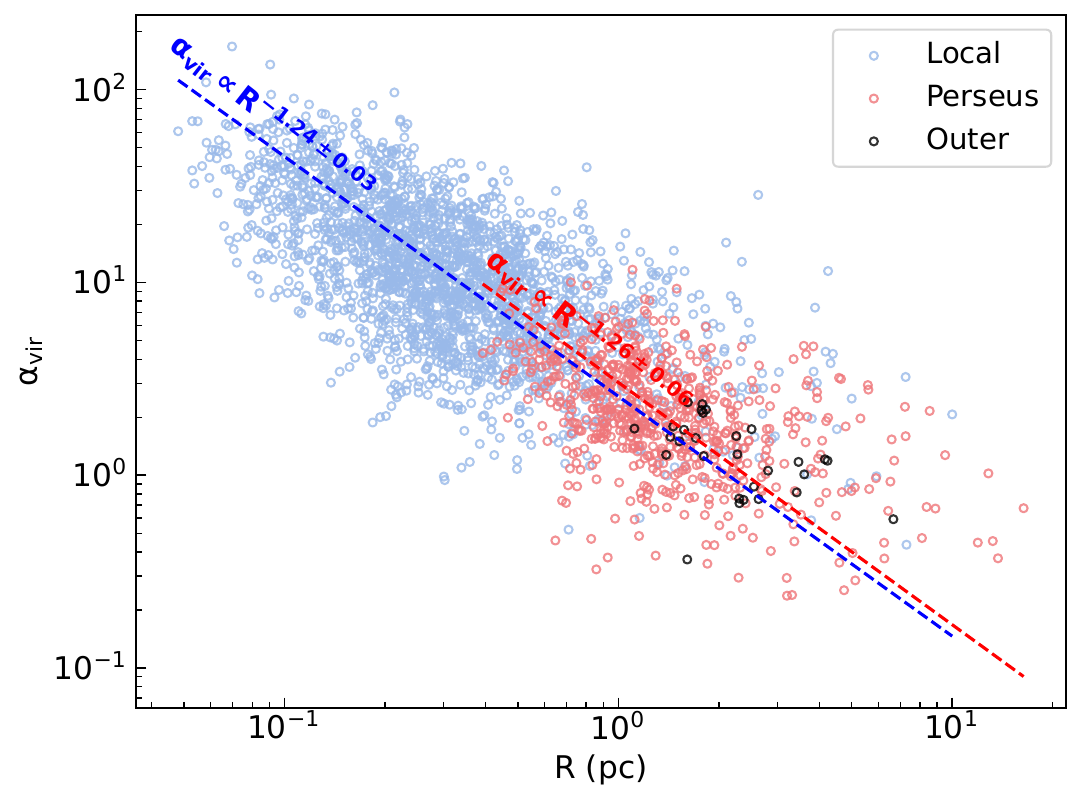}
		\\(a)
	\end{minipage}
	\begin{minipage}[t]{0.315\linewidth}
		\centering
		\includegraphics[width=\linewidth]{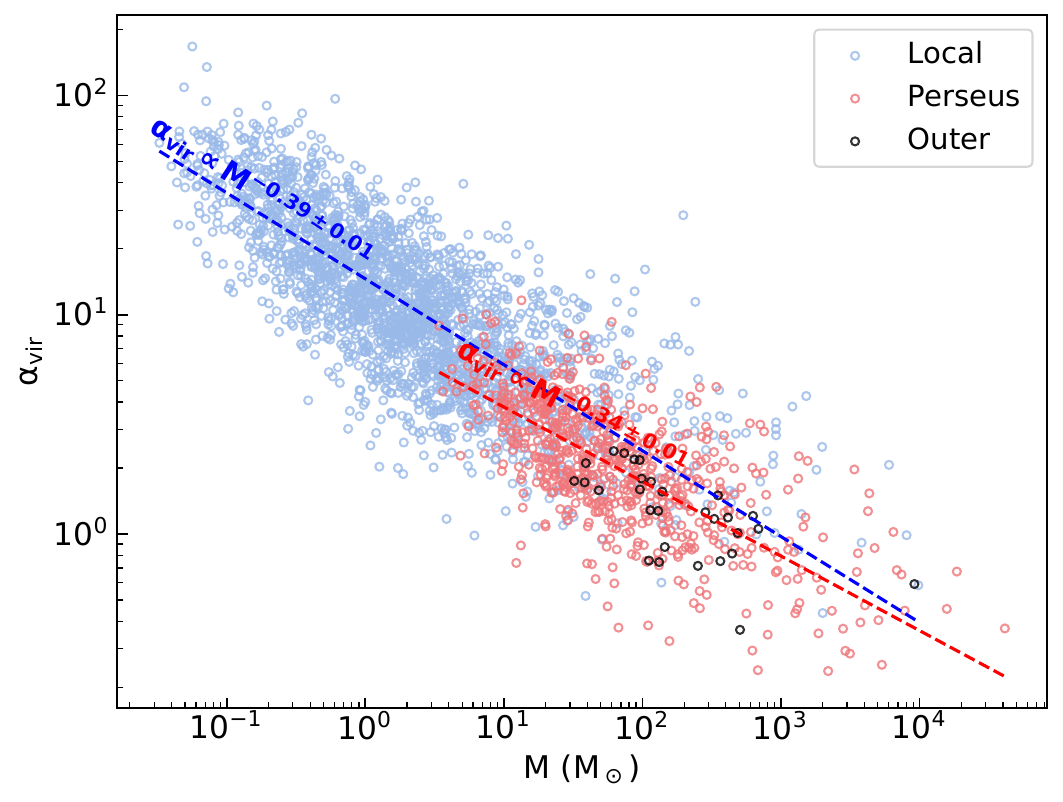}
		\\(b)
	\end{minipage}
	\begin{minipage}[t]{0.32\linewidth}
		\centering
		\includegraphics[width=\linewidth]{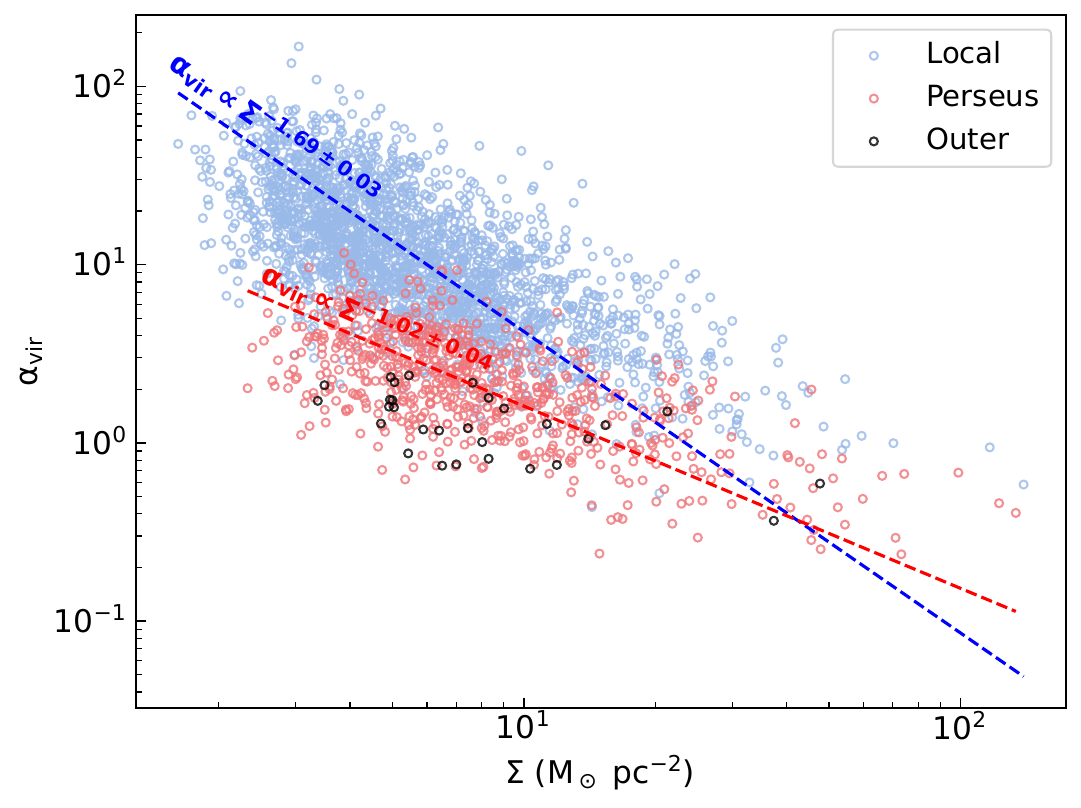}
		\\(c)
	\end{minipage}
	\caption{Variation of virial parameter as (a) radius, (b) mass, and (c) surface density for $^{13}$CO clouds. Structures in different spiral arms are indicated with different color. The blue dashed lines and texts indicate the fitting result for molecular clouds in the Local Arm, while the red ones present those for molecular clouds in the Perseus and Outer Arms.}\label{fig:variation of vp}
\end{figure*}

Figure \ref{fig:variation of vp}(a) plots the virial parameter against radius, showing that a larger  radius is associated with a lower virial parameter. In the Local Arm, molecular clouds yield the relation of $\alpha_\mathrm{vir} \propto R^{-1.24\pm 0.03}$, while in the Perseus and Outer Arms, the power-law relations become slightly steeper with an index of $-1.26 \pm 0.06$. The fitted power-law indices, which remain relatively constant across different spiral arms, are slightly steeper than those proposed by \cite{Krumholz2019}. 

The virial parameter of molecular clouds as a function of mass is presented in Figure \ref{fig:variation of vp}(b), giving the relation of $\alpha_\mathrm{vir} \propto M^{-0.39 \pm 0.01}$ in the Local Arm, and $\alpha_\mathrm{vir} \propto M^{-0.34 \pm 0.01}$ in the Perseus and Outer Arms. The power-law indices we derived are slightly shallower than the value of $-0.53\pm0.30$ reported by \cite{Miville2017} for their $^{12}$CO clouds. The $\alpha_\mathrm{vir}-M$ relation shows a trend similar to the $\alpha_\mathrm{vir}-R$ relation, only with smaller power-law indices. It is worth noting that \cite{Chevance2023} have pointed out that a relation of $\alpha_\mathrm{vir} \propto M^{0.5}$ is expected for the minimum detectable virial parameter at a fixed surface brightness sensitivity and velocity resolution. Observational limitations and sample selection effects may also influence the $\alpha_\mathrm{vir}-M$ relation. For example, we may not be able to identify self-gravitating clouds with low masses, as these clouds also have very low velocity dispersions.

Figure \ref{fig:variation of vp}(c) illustrates the variation of the virial parameter with surface density, revealing a negative correlation. The power-law indices vary significantly in different spiral arms. Molecular clouds in the Local Arm yield the relation of $\alpha_\mathrm{vir} \propto \Sigma^{-1.69 \pm 0.03}$, and the power-law index is $-1.02\pm 0.04$ for molecular clouds in the Perseus and Outer Arms. Although there are no significant systematic differences in the surface density of molecular clouds across different arms, molecular clouds in the Local Arm tend to have a higher virial parameter than those in the Perseus and Outer Arms at a given surface density, possibly indicating stronger turbulence in the Local Arm. As can be inferred from the definition of virial parameter, distance measurements influence the estimates of molecular cloud scales and masses, but the derivation of velocity dispersion is independent of distance, resulting in a relation of $\alpha_\mathrm{vir}\propto D^{-1}$. As a consequence, the systematically higher $\alpha_\mathrm{vir}$ of molecular clouds in the Local Arm may result from an underestimate of their kinematic distances.


\subsection{Single-cloud Analysis} \label{sec:3.2}

\subsubsection{Scaling Relations}\label{subsec:3.2.1}
In this section, we study scaling relations among substructures within a single cloud, which we refer to as type 4 scaling relations.
To ensure statistically robust correlation analysis, we selected molecular clouds containing more than 15 substructures to be our sample, resulting in a total of 50 qualified molecular clouds. For each cloud, we analyze four scaling relations among its substructures, namely those between $\sigma_v$ and $R$, $M$ and $R$, $\mathcal{L}$ and $\Sigma$, and $\sigma_v$ and $\Sigma R$.

\begin{figure*}[!htb]
	\centering
	\includegraphics[width=\textwidth]{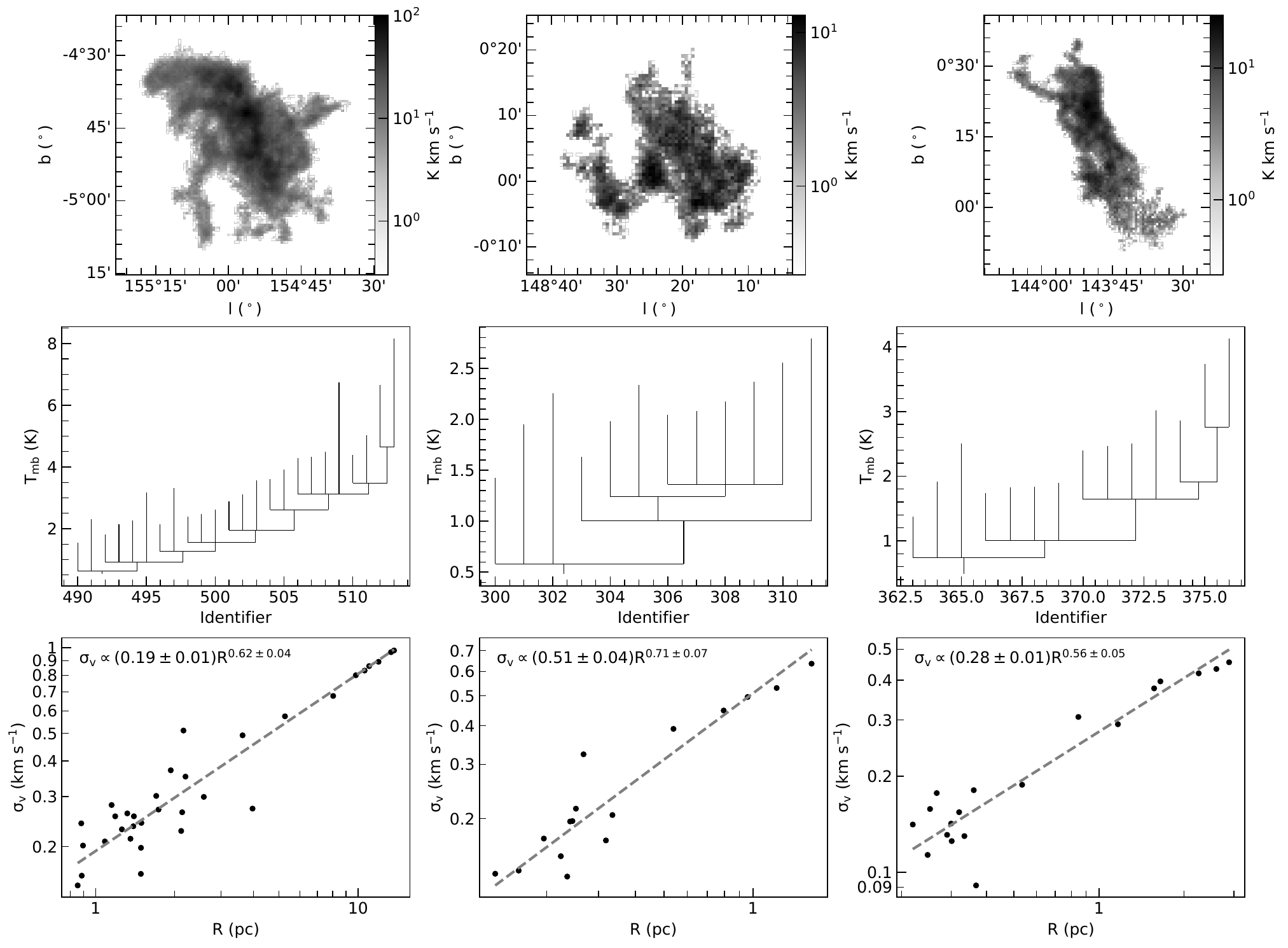}
	\caption{Top: $^{13}$CO integrated intensity maps of three individual molecular clouds as examples. 
Middle: The tree diagrams of the molecular clouds.
Bottom: Linewidth-size relation of the substructures. Gray dashed lines denote the fitting results.}\label{fig:cases}
\end{figure*}

Figure \ref{fig:cases} displays the integrated intensity maps, tree diagrams, and corresponding type 4 linewidth-size relations of three $^{13}$CO clouds and their substructures as examples. Both the power-law indices and constant terms of the linewidth-size relation vary significantly across different molecular clouds, highlighting the necessity of investigating type 4 scaling relations across a diverse sample of molecular clouds. Further comparisons beween the type 2 and the type 4 scaling relations are given in Section \ref{subsec:4.1}.

\begin{figure*}[!htb]
	\centering
    \begin{minipage}[t]{0.49\linewidth}
        \centering
	    \includegraphics[width=\textwidth]{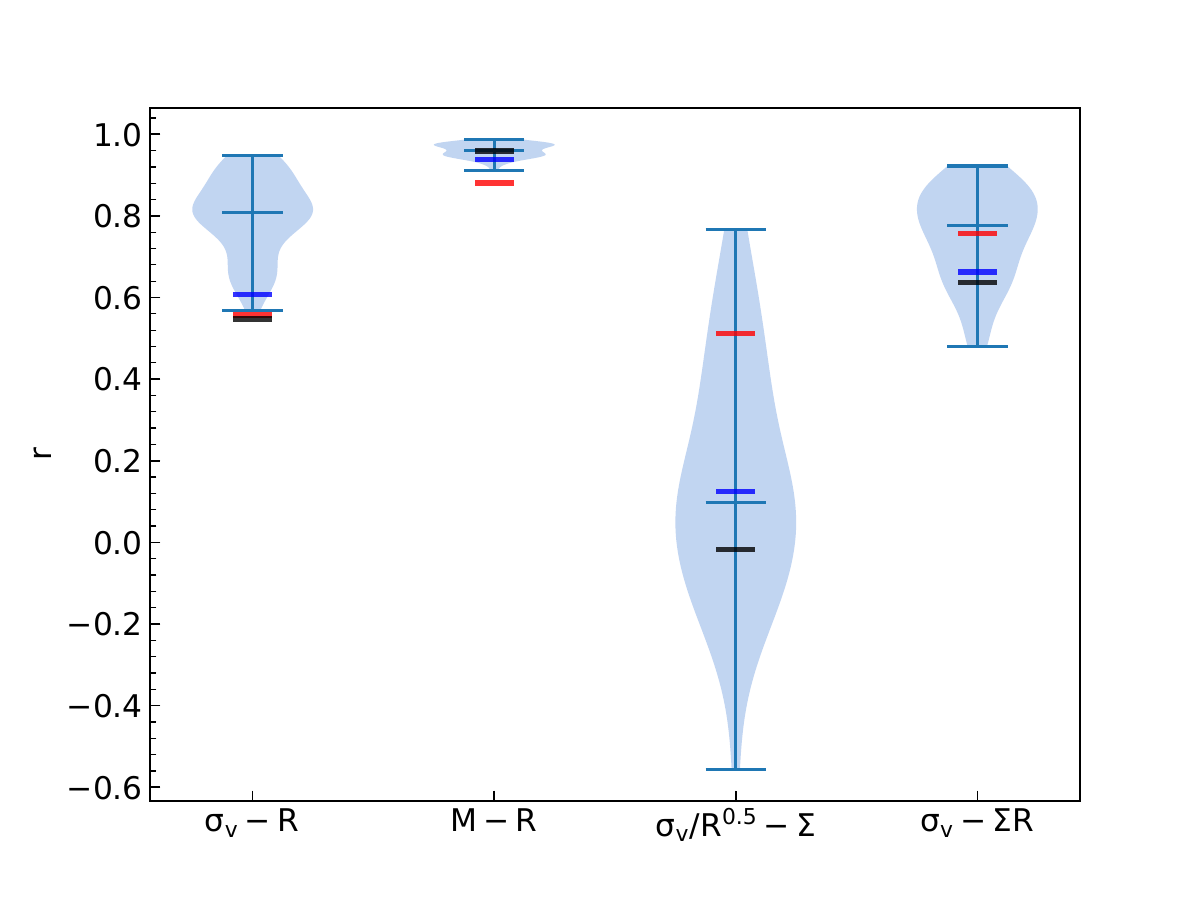}
        \\(a)
    \end{minipage}
    \begin{minipage}[t]{0.49\linewidth}
        \centering
	    \includegraphics[width=\textwidth]{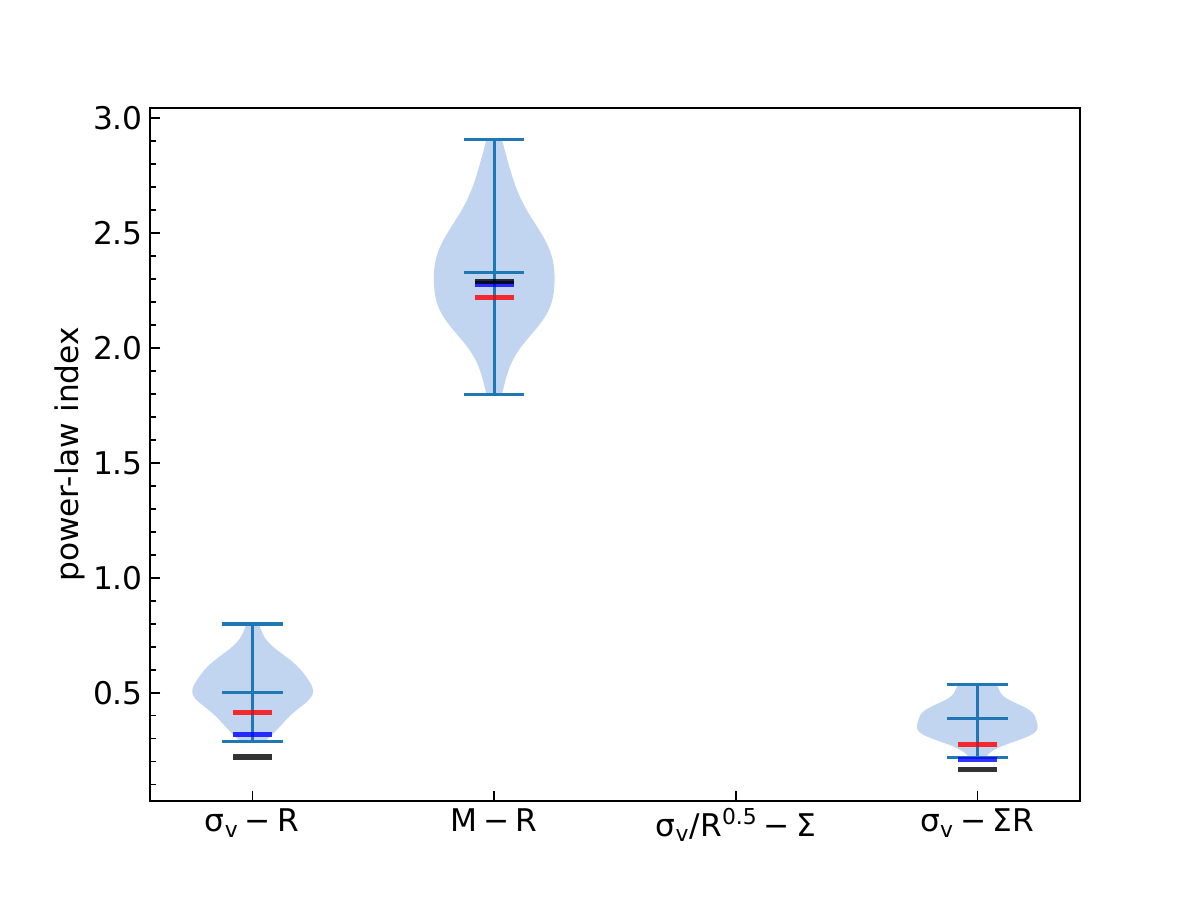}
        \\(b)
    \end{minipage}
	\caption{(a) Violin plots of Pearson correlation coefficients between the logarithms of $\sigma_v$ and $R$, $M$ and $R$, $\mathcal{L}$ and $\Sigma$, and $\sigma_v$ and $\Sigma R$ of hierarchical structures within molecular clouds with more than 15 substructures. (b) Violin plots of the fitted power-law indices of the type 4 linewidth-size relation, mass-size relation, revised Keto-Heyer relation, and the relation between $\sigma_v$ and $\Sigma R$ within these molecular clouds. Only fitting results with a Pearson correlation coefficient greater than 0.75 are included. The Keto-Heyer relation is excluded due to the instability of its Pearson correlation coefficient. The three light blue horizontal lines from top to bottom represent the maximum, median, and minimum values of the distribution, respectively. The blue, red, and black short horizontal lines represent the type 2 Pearson correlation coefficients and the power-law exponents obtained from fits to molecular clouds belonging to the Local Arm, molecular clouds belonging to the Perseus and Outer Arms, and all molecular clouds, respectively.}\label{fig:pearson}
\end{figure*}

Figure \ref{fig:pearson}(a) presents the violin plots of the Pearson correlation coefficients between the physical parameters involved in the scaling relations. The mass-size relation exhibits the strongest correlation among all molecular clouds, with a median Pearson correlation coefficient of 0.96 and a minimum value of 0.91. 
In the case of type 2, the $\sigma_v - R$ correlation is slightly weaker than that between $\sigma_v$ and $\Sigma R$, while for type 4, the $\sigma_v - R$ correlation is stronger. The maximum, median, and minimum Pearson correlation coefficients between $\log{\sigma_v}$ and $\log{R}$ are 0.95, 0.81, and 0.57 respectively, all higher than the corresponding values, which are 0.92, 0.78, and 0.48 for the $\sigma_v - \Sigma R$ relation. However, the correlation between $\mathcal{L}$ and $\Sigma$ varies significantly among molecular clouds, with Pearson correlation coefficients ranging from -0.56 to 0.76 and a median value of 0.10.
As illustrated by \cite{Vazquez2025}, for molecular clouds whose linewidth-size relation originates from the Burgers-like turbulent energy spectrum ($E_k \propto k^{-2}$) expected for strongly supersonic turbulence, $\mathcal{L}$ should remain constant for all substructures inside it. Conversely, if the velocity dispersion is dominated by gravitational contraction motions as proposed in GHC, then the Keto-Heyer relation $\mathcal{L}\propto \Sigma^{0.5}$ is expected. 
The relatively strong correlations found for the type 4 linewidth-size relation, together with the lack of consistently high Pearson correlation coefficients in the Keto-Heyer relation, may indicate that in $^{13}$CO-traced molecular gas, turbulence plays a more dominant role in shaping the velocity dispersion at sub-cloud scales than predicted by the GHC model.
Alternatively, this may reflect the fact that, for hierarchical structures identified as nested entities in PPV space, the projected surface density may not reliably trace their structural compactness. We return to this issue in Section \ref{subsubsec:KH}, where we examine a revised Keto--Heyer relation formulated in terms of volume density.

Figure \ref{fig:pearson}(b) shows the fitted power-law indices of the type 4 scaling relations within molecular clouds with more than 15 substructures. The black horizontal lines show the type 2 power-law indices derived from fitting all molecular clouds, while the blue and red lines represent the results for molecular clouds in the Local Arm and the Perseus and Outer Arms, respectively.

The power-law indices of the type 4 linewidth-size relation range from 0.26 to 0.80. The median value (0.50) aligns well with the prediction of the \textit{Burgers'} turbulence. The minimum, median, and maximum power-law indices of the type 4 mass-size relation are 1.80, 2.33, and 2.91, respectively. The power-law index of the mass-radius relation may be influenced by factors such as the morphology of molecular clouds and their substructures, their density distributions, and their inclination angles relative to the line of sight. The Keto-Heyer relation is excluded due to the large fluctuations observed in its Pearson correlation coefficient. Finally, for the relation between $\sigma_v$ and $\Sigma R$, its power-law indices show the smallest range, spanning from a minimum of 0.22 to a maximum of 0.53, with a median value of 0.39. Our result, obtained using the type 4 approach, is consistent with $\sigma_v \propto (\Sigma R)^{0.43\pm0.14}$ reported by \citet{Miville2017} in a sample of Galactic $^{12}$CO clouds.

The scenario in which gravity is dominant claims that instead of linewidth-size relation, the more general underlying relation would be that of energy equipartition \citep{Vazquez2025}, which implies $|E_g|\sim E_k$ so that $\sigma_v \sim \sqrt{GM/R} \sim \sqrt{\pi G\Sigma R}$. As a result, the Larson linewidth-size relation is only expected to hold for gravitationally-dominated objects of approximately the same column density \citep{BP2011b}. Instead, when considering objects of a wide range of column densities, one should expect the Keto-Heyer relation, $\sigma_v/R^{0.5}\sim \Sigma$, to hold.

As shown in Figure \ref{fig:pearson}(a), the Pearson correlation coefficient of the Keto-Heyer relation in the classical form of $\sigma_v/R^{0.5}$ versus $\Sigma$ for substructures within individual molecular clouds is not high. However, we argue that this does not directly imply that the substructures of molecular clouds are inconsistent with the predictions of the GHC model. Instead, it is because for nested structures identified using the Dendrogram algorithm in PPV space, the surface density is no longer a reliable measure of structural compactness. As a result, we propose a revised Keto-Heyer relation, i.e., the relationship between $\sigma_v/R$ and $\rho$, in which $\rho$ is the volume density of the structure. A more detailed discussion is presented in Section \ref{subsubsec:KH}. 

\begin{figure*}[!htb]
	\centering
	\begin{minipage}[t]{0.49\linewidth}
		\centering
		\includegraphics[width=\linewidth]{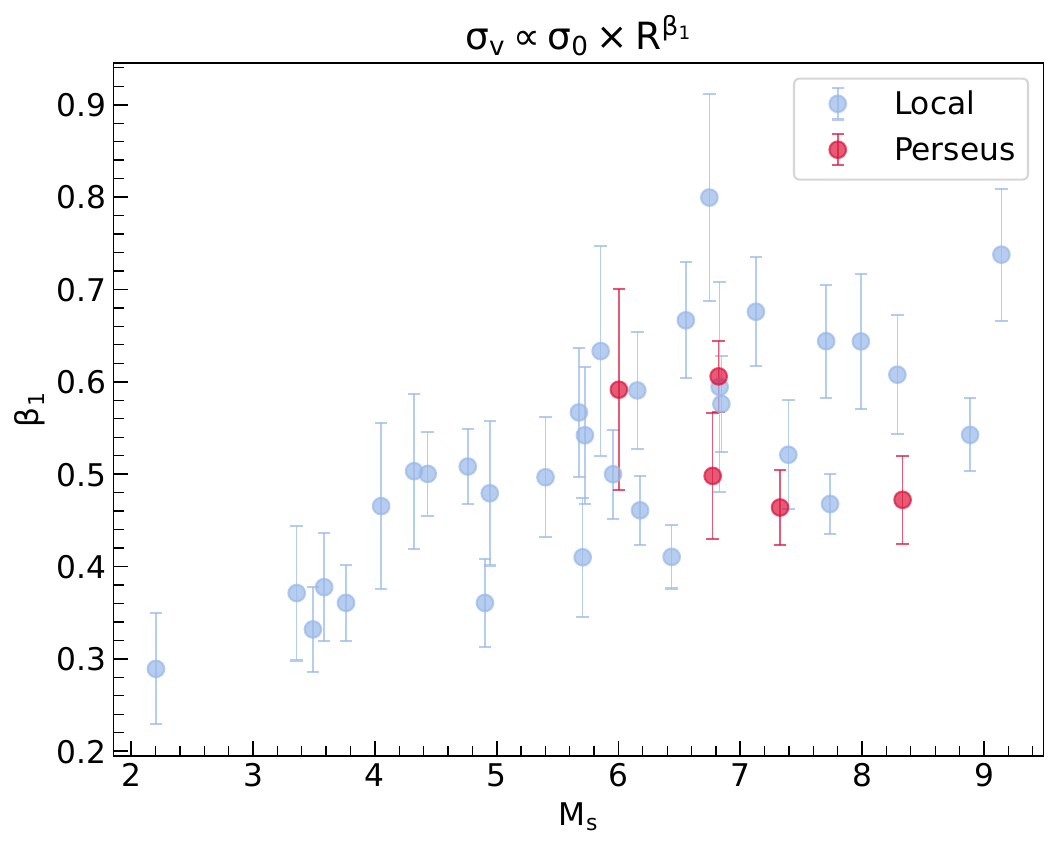}
		\\(a)
	\end{minipage}
	\begin{minipage}[t]{0.49\linewidth}
		\centering
		\includegraphics[width=\linewidth]{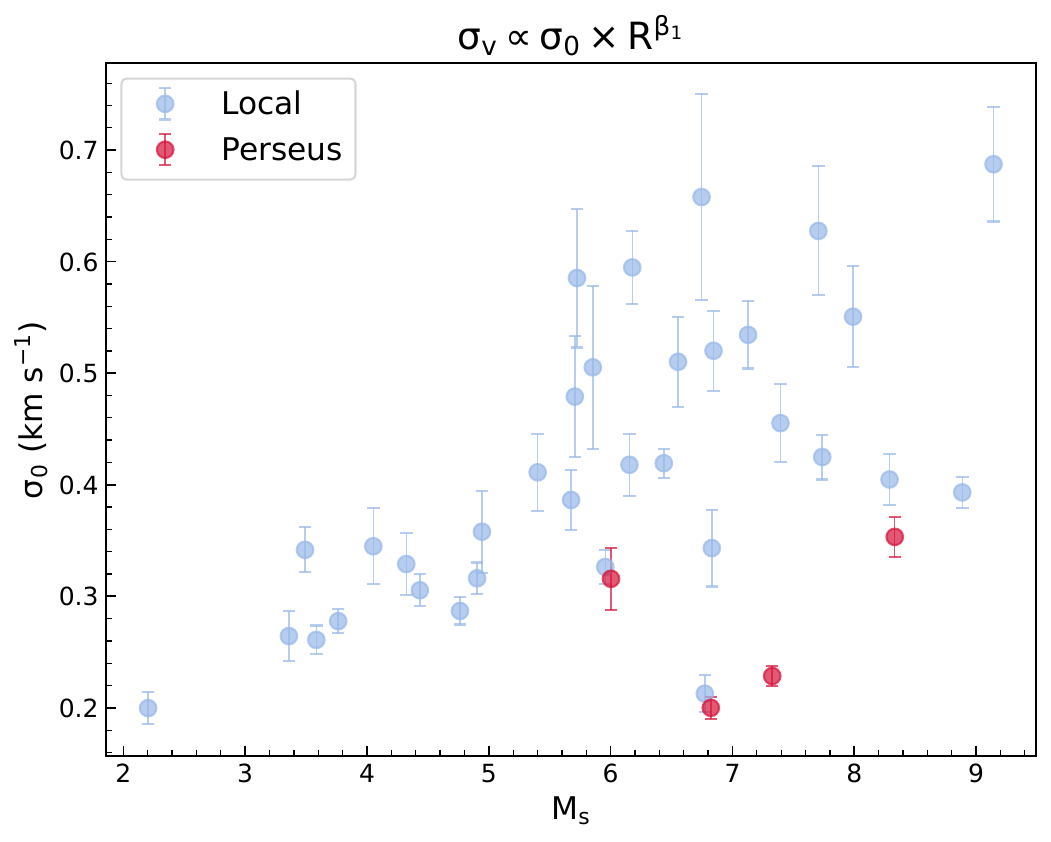}
		\\(b)
	\end{minipage}
	\begin{minipage}[t]{0.49\linewidth}
		\centering
		\includegraphics[width=\linewidth]{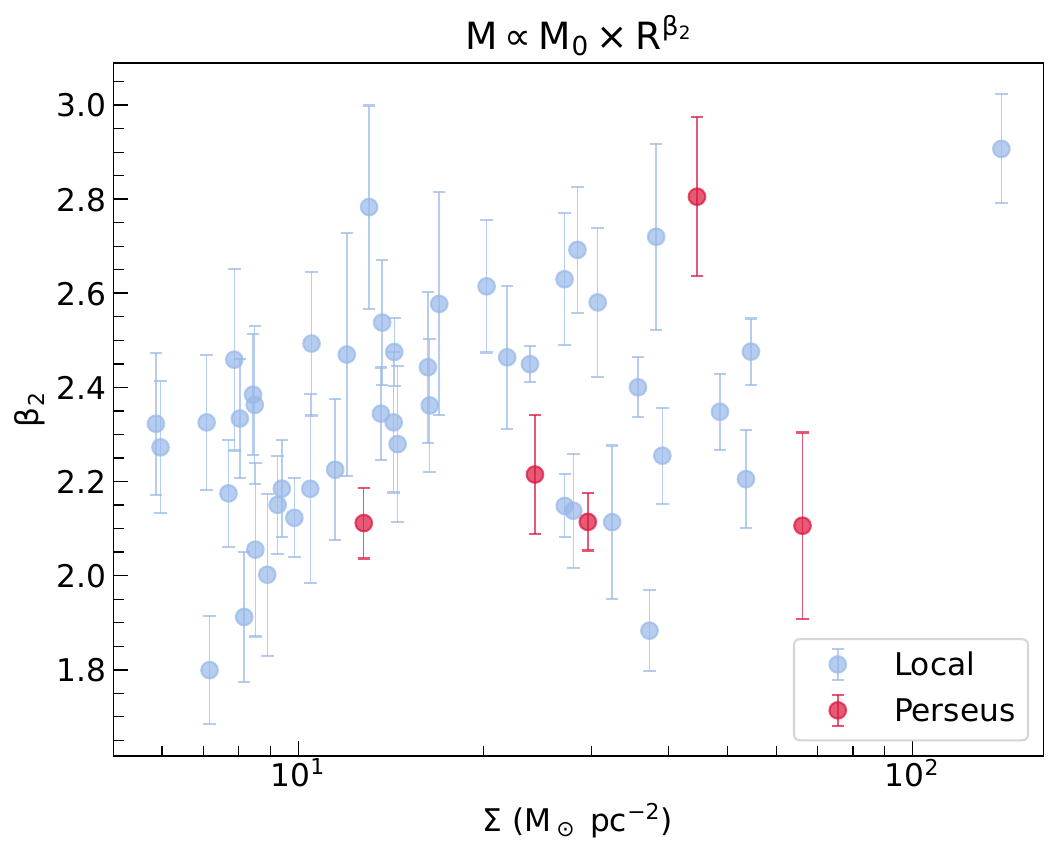}
		\\(c)
	\end{minipage}
	\begin{minipage}[t]{0.49\linewidth}
		\centering
		\includegraphics[width=\linewidth]{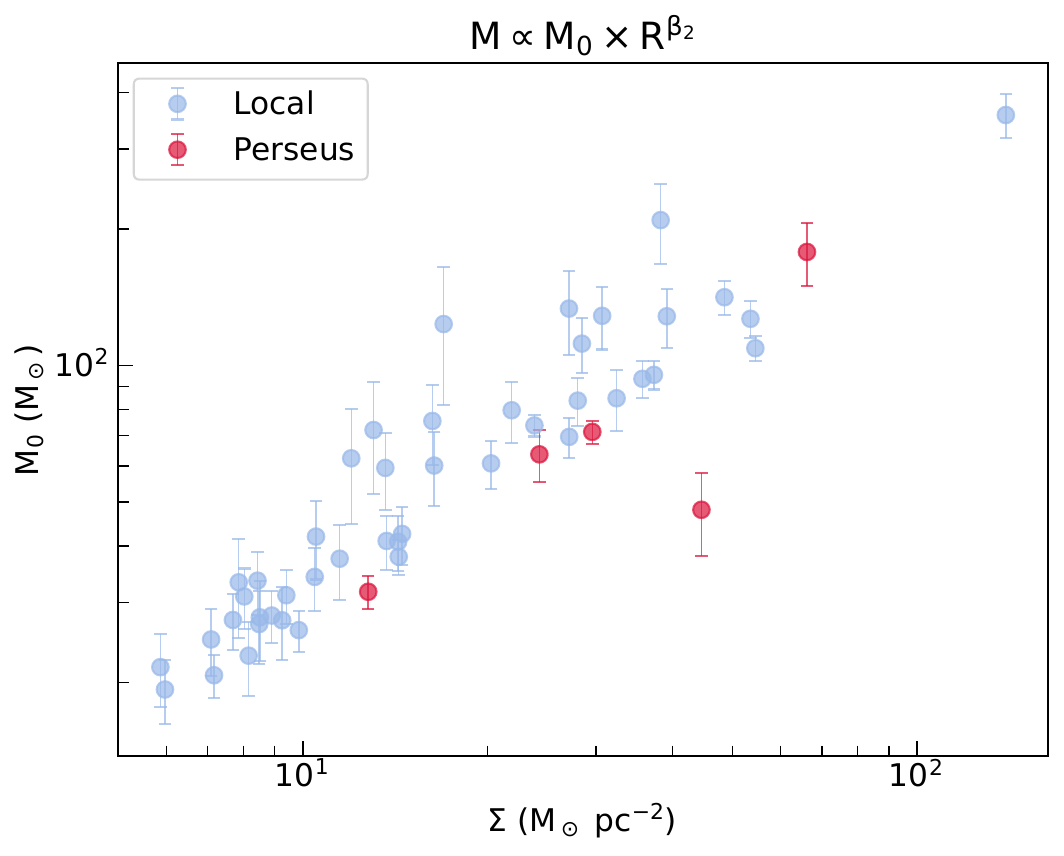}
		\\(d)
	\end{minipage}
	\caption{(a) The relation between the power-law exponent of the $\sigma_v - R$ relation derived from hierarchical substructures identified by the non-binary Dendrogram algorithm from each molecular cloud and the sonic Mach number of the cloud. (b) Same as (a) but the y-axis stands for the constant term instead. (c) Same as (a) but for the power-law index of $M-R$ relation and the surface density of the cloud. (d) Same as (c) but the y-axis stands for the constant term of the fitting. Only the fitting results with a Pearson correlation coefficient greater than 0.75 are counted.}\label{fig:beta_Ms}
\end{figure*}

We further test whether the forms of the type 4 linewidth-size relation and the mass-size relation are universal. To minimize potential biases introduced by fitting procedures, we restricted our analysis to type 4 scaling relations with Pearson correlation coefficients greater than 0.75. The top two panels of Figure \ref{fig:beta_Ms} illustrate the variation of the power-law index, $\beta_1$, and the constant term, $\sigma_0$, of the $\sigma_v-R$ relation with the sonic Mach number, $M_\mathrm{s}$, of the molecular cloud. As shown in Figure \ref{fig:beta_Ms}(a), instead of stabilizing around the canonical value of 0.5, the $\beta_1$ of the type 4 linewidth-size relation shows a strong positive correlation with the $M_\mathrm{s}$ of the molecular cloud and ranges from around 0.3 to 0.8, and a similar but less robust trend is obtained from the relation between $\sigma_0$ and $M_\mathrm{s}$ in Figure \ref{fig:beta_Ms}(b). In addition to thermal broadening, the derivation of the velocity dispersion for substructures is also influenced by large-scale velocity gradients, and we further discuss the impact of removing cloud-scale velocity gradients on the results in Appendix~\ref{appA1}. 

Figure \ref{fig:beta_Ms}(c) and (d) show how the power-law index, $\beta_2$, and the constant term, $M_0$ of the type 4 $M-R$ relation vary with the surface density, $\Sigma$, of the cloud. Since $M_0$ is essentially the fitted mass $M$ when $R$ is normalized to unity, it equivalently represents a characteristic surface density. The $M_0$ exhibits a strong correlation with the surface density of molecular clouds, while the $\beta_2$ shows no correlation with $\Sigma$ and fluctuates around the characteristic value of $2.2-2.3$ for the type 2 mass-size relation. The variation of the power-law indices of type 4 $M-R$ relation may be closely related to the column density profiles of molecular clouds as proposed by \cite{Xing2022}.

The power-law exponent $\beta_1$ of the linewidth-size relation, which deviates from 0.5, has also been observed in previous studies of potentially star-forming clumps. For instance, \citet{Traficante2020} investigated a sample of $70-\mathrm{\mu m}$ quiet clumps and their parent filaments, finding that clumps with high mass surface density ($\rm \Sigma > 0.1\ g\ cm^{-2}$) exhibit a flatter $\beta_1$ compared to those with lower surface density. This is attributed to the dynamics of high-surface-density clumps being predominantly governed by gravitational collapse. \citet{Peretto2023} studied the velocity dispersion profiles across 27 infrared dark molecular clouds and their parent molecular clouds and revealed that clump-scale structures traced by $\rm N_2H^+ (1-0)$ emission display flat velocity dispersion profiles, while the $\beta_1$ values derived with $\rm ^{13}CO(1-0)$ emission for cloud-scale velocity dispersion profiles show significant scatter. However, to the best of our knowledge, the correlation between $\beta_1$ and $M_\mathrm{s}$ has not been previously reported. 

The linewidth-size relation is commonly interpreted as a manifestation of the turbulent energy cascade, with different power-law exponents reflecting different types of turbulence. For instance, Kolmogorov's incompressible turbulence corresponds to $\beta_1 = 1/3$, while a Burgers-like turbulent energy spectrum corresponds to $\beta_1 = 1/2$. Therefore, the observed correlation between $\beta_1$ and $M_\mathrm{s}$ suggests either that the velocity scaling properties of turbulence in molecular clouds depend on the level of compressibility (as measured by $M_\mathrm{s}$), or that the linewidth-size relation cannot be interpreted purely as a manifestation of a scale-free turbulent cascade.

\subsubsection{Variation of Virial Parameter} \label{sec:3.2.2}
\begin{figure*}[!htb]
	\centering
    \begin{minipage}[t]{0.49\linewidth}
        \centering
        \includegraphics[width=\textwidth]{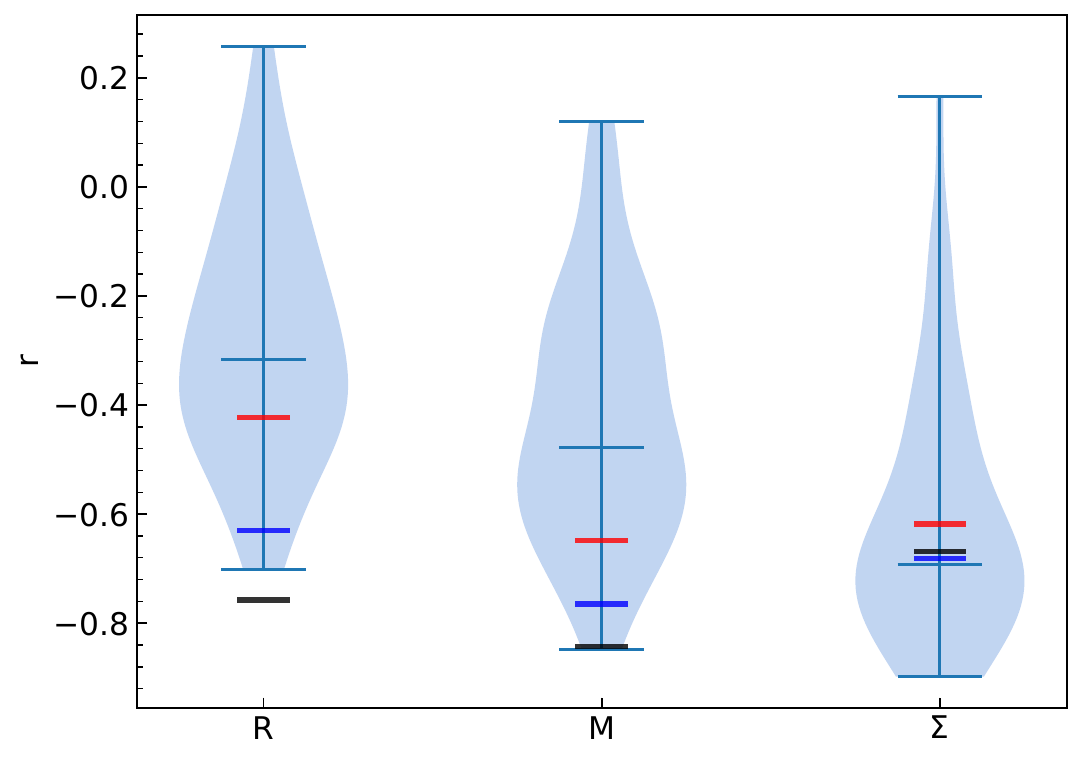}
        \\(a)
    \end{minipage}
    \begin{minipage}[t]{0.49\linewidth}
        \centering
        \includegraphics[width=\textwidth]{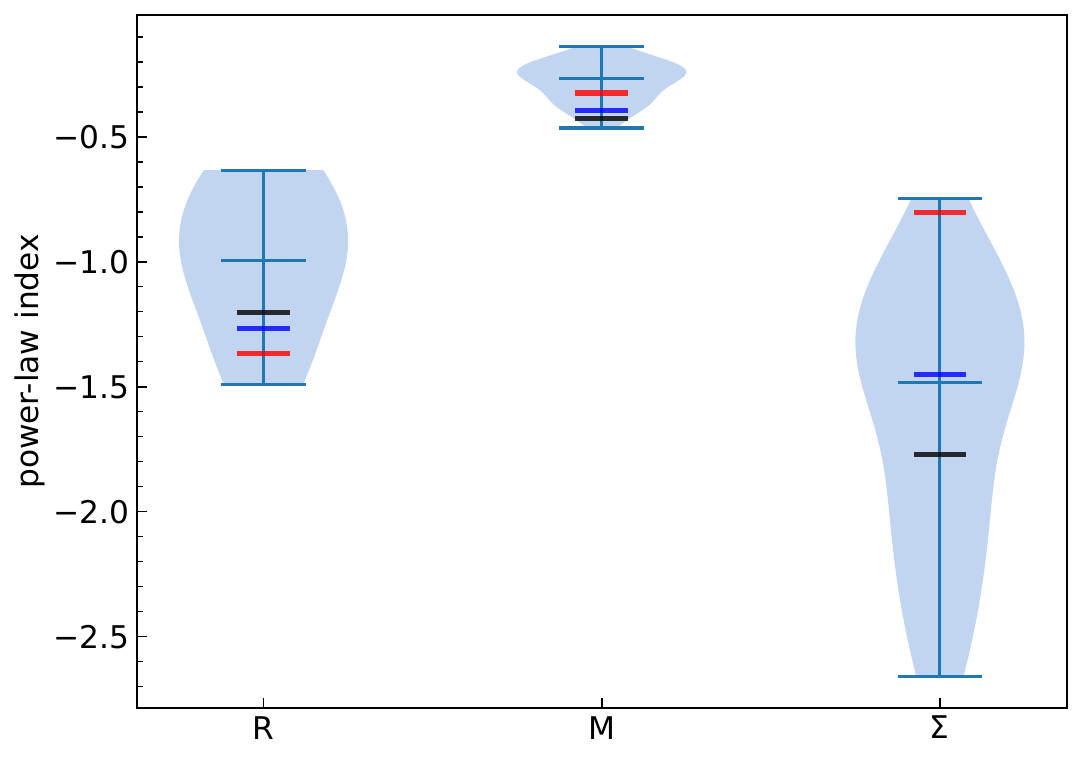}
        \\(b)
    \end{minipage}
	
	\caption{(a) Violin plots of Pearson correlation coefficients between the logarithms of the virial parameter and radius, mass, and surface density of hierarchical structures within molecular clouds with more than 15 substructures. (b) Same as (a) with the y-axis representing the power-law indices instead, only fitting results with an absolute Pearson correlation coefficient above 0.5 are retained. The three horizontal light blue lines from top to bottom correspond to the maximum, median, and minimum of the distribution, respectively. The blue, red, and black short lines indicate the type 2 Pearson correlation coefficients and the power-law indices for molecular clouds belonging to the Local Arm, molecular clouds belonging to the Perseus and Outer Arms, and all molecular clouds, respectively.}\label{fig:vp_param}
\end{figure*}

It has been widely observed that the virial parameter of molecular clouds is negatively correlated with their size and mass, suggesting that larger and heavier molecular clouds are more likely to be gravitationally bound. Here we further investigate whether such trends still apply to the substructures of molecular clouds. Figure \ref{fig:vp_param}(a) shows the correlation between the virial parameter and several other physical properties of the hierarchical substructures within molecular clouds. In general, the virial parameter is anti-correlated with radius, mass, and surface density of the structure.
The virial parameter is most strongly influenced by surface density, with a median Pearson correlation coefficient of -0.69, while the correlation with radius and mass is slightly weaker, with median Pearson correlation coefficients of -0.32 and -0.48, respectively. The relatively low correlations of $\alpha_\mathrm{vir}$ with both $M$ and $R$, compared to the type 2 case, may suggest that the dynamics at sub-cloud scales are far more complex than those at the cloud scale.

Figure \ref{fig:vp_param}(b) presents violin plots of the power-law indices for the relations between the virial parameter and radius, mass, and surface density of substructures within molecular clouds. The corresponding maximum, median, and minimum values for the $\alpha_\mathrm{vir}-R$ relation are -1.49, -1.00, and -0.63, respectively. The distribution is tightest for the $\alpha_\mathrm{vir}-M$ relation, with values of -0.46, -0.26, and -0.14. Despite showing the strongest correlation, the $\alpha_\mathrm{vir}-\Sigma$ relation exhibits the largest scatter in its power-law index, with values spanning from -2.66 to -0.75 and a median of -1.48. Finally, we note that the power-law indices obtained on a cloud-by-cloud basis across the three spiral arms all fall within the distribution derived from substructures.

Based on the Pearson correlation coefficients, the majority of clouds in our sample would be classified as exhibiting a “flat/falling” virial profile—i.e., the virial parameter is either uncorrelated with or negatively correlated with surface density—under the differential virial analysis framework of \citet{Krumholz2025}, who associate such profiles with gravitational contraction. However, it should be noted that our methodology differs from theirs, particularly in terms of structure identification using three-dimensional PPV brightness temperature isosurfaces rather than a two-dimensional column density threshold in their work.

\section{Discussion}\label{sec:4}
\subsection{Comparison between the Type 2 and Type 4 Scaling Relations}\label{subsec:4.1}
In this section, we compare the similarities and differences between type 2 and type 4 scaling relations. In Figure \ref{fig:pearson}, the black, blue, and red horizontal lines show the type 2 Pearson correlation coefficients and power-law indices derived from fitting molecular clouds associated with different spiral arms. The linewidth-size relation exhibits stronger correlations among substructures within individual molecular clouds than among the molecular clouds themselves, as the type 2 Pearson correlation coefficients are around 0.6, while the median type 4 value is 0.81. The generally lower type 2 Pearson coefficients compared to type 4, along with the deviation of type 2 power-law indices from the canonical value of 0.5, suggest that multicloud scaling relations are affected by cloud-to-cloud variations in physical conditions and observational selection effects. In contrast, type 4 scaling relations compare substructures within the same molecular cloud, and therefore provide a more direct probe of the internal dynamical organization of individual clouds.

Although the type 2 mass-size relation already exhibits a high correlation coefficient, it still falls below the median correlation coefficient of the type 4 mass-size relation. The power-law indices of the type 2 mass-size relation are generally higher than the median value of the type 4 mass-size relation, yet still lie within the distribution of type 4 power-law indices. 

When considering the Keto-Heyer relation, it exhibits relatively high correlations across the molecular cloud sample, whereas no stable correlation is observed under the type 4 scheme. However, as we will see in Section \ref{subsubsec:KH}, when we replace the surface density in the Keto-Heyer relation with volume density, its correlation shows a significant improvement.

The Pearson correlation coefficients of the relation between $\sigma_v$ and $\Sigma R$ are slightly higher in the type 4 scheme, as the three type 2 Pearson coefficients are all smaller than the median type 4 value. The type 2 power-law indices of the relation between $\sigma_v$ and $\Sigma R$ derived from molecular clouds in different arms are all shallower than the median value in the type 4 situation (0.39). 

Overall, type 4 scaling relations consistently exhibit stronger correlations, except for the Keto-Heyer relation where no stable correlation is observed in the type 4 case. In general, the power-law indices of the type 2 scaling relations fall within the range of those for the type 4 scaling relations, except for the linewidth-size relation and the relation between $\sigma_v$ and $\Sigma R$ obtained by fitting all molecular clouds. The above consistences and differences between the two schemes suggest that the scaling relations derived from cloud populations do not necessarily represent the intrinsic dynamical behavior within individual molecular clouds. While the type 2 relations capture the overall statistical trends across the surveyed region, their slopes and correlation strengths are likely affected by cloud-to-cloud variations in surface density, gravitational state, and global Mach number.

By contrast, the tighter correlations obtained in the type 4 scheme imply that velocity and mass scaling are more coherent within a common hierarchical system. It should be noted that the type 4 scaling relations involve repeated use of some emission voxels due to the hierarchical nesting of Dendrogram structures, which may partly affect the measured correlations. To assess this effect, in Appendix~\ref{appB} we perform an additional test by randomly selecting structures with no overlapping voxels within each molecular cloud and re-evaluating the correlations. The resulting correlation coefficients remain systematically higher than those obtained for the Type 2 relations, indicating that the enhanced correlations in the Type 4 scheme cannot be attributed solely to voxel overlap.
The absence of a stable Keto–Heyer relation at the internal level further suggests that virial-like scaling does not hold across all substructures of a given cloud. Instead, different hierarchical levels may occupy distinct dynamical states, reflecting the combined influence of turbulence and gravity rather than a single universal equilibrium condition. In this sense, the type 4 relations provide additional insight into the internal dynamical organization of molecular clouds beyond what can be inferred from multicloud statistics alone.

\subsection{The Revised Keto-Heyer Relation}\label{subsubsec:KH}

Based on the proposal by \cite{Vazquez2025}, a key prediction of the GHC model is that molecular clouds and their substructures in a state of near free-fall should exhibit a state analogous to energy conservation. This is manifested by their distribution near the line where the virial parameter equals 2 in the Keto-Heyer diagram. The Keto-Heyer diagram, where the horizontal axis represents the surface density, $\Sigma$, of the structure, and the vertical axis represents the Larson ratio, $\mathcal{L}\equiv\sigma_v/R^{0.5}$, is directly related to the definition of the virial parameter:
\begin{equation}
    \alpha_\mathrm{vir}\equiv \frac{5\sigma_v^2R}{GM}=\frac{5\sigma_v^2/R}{\pi G\Sigma} = \frac{5\mathcal{L}^2}{\pi G\Sigma}.
\end{equation}
When molecular clouds are treated as discrete units, their projected spatial scale can be determined from a density or intensity threshold. Similarly, the velocity range of the cloud can be reasonably constrained, allowing its total flux to be summed and its mass to be derived. As a result, surface density is commonly used to describe the compactness of molecular clouds. However, for hierarchical substructures within one molecular cloud identified by methods like Dendrogram, volume density may be a more appropriate descriptor. Note that under the assumption of an approximately spherical geometry,
\begin{equation}
    \alpha_\mathrm{vir}\equiv\frac{5\sigma_v^2R}{GM}=\frac{5\sigma_v^2/R}{\pi G\Sigma}=\frac{15\sigma_v^2/R^2}{4\pi G\rho},
\end{equation}
indicating that the relation of $\mathcal{L}$ versus $\Sigma$ should be equivalent to that of $\sigma_v/R$ versus $\rho$. 

\begin{figure*}[!htb]
	\centering
    \begin{minipage}[t]{0.49\linewidth}
        \centering
	    \includegraphics[width=\textwidth]{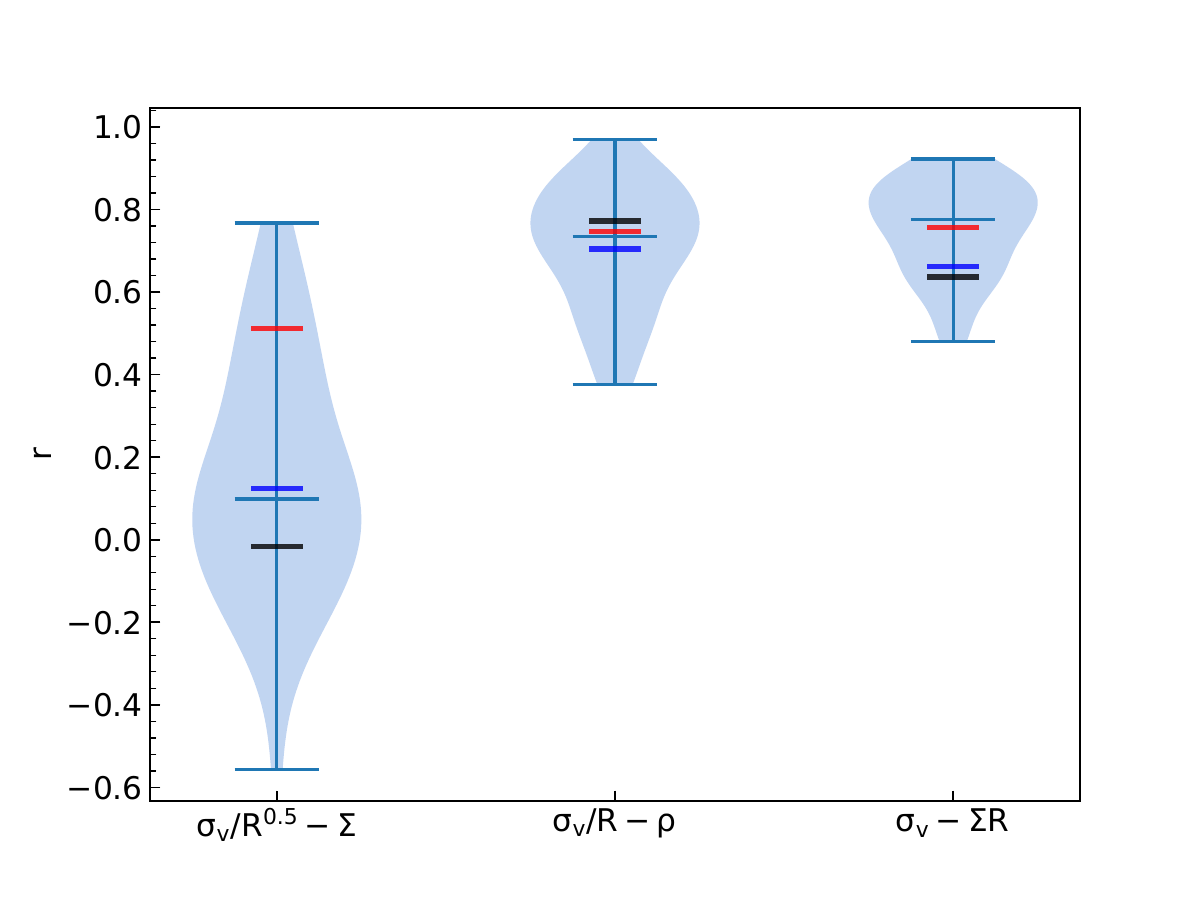}
        \\(a)
    \end{minipage}
    \begin{minipage}[t]{0.49\linewidth}
        \centering
	    \includegraphics[width=\textwidth]{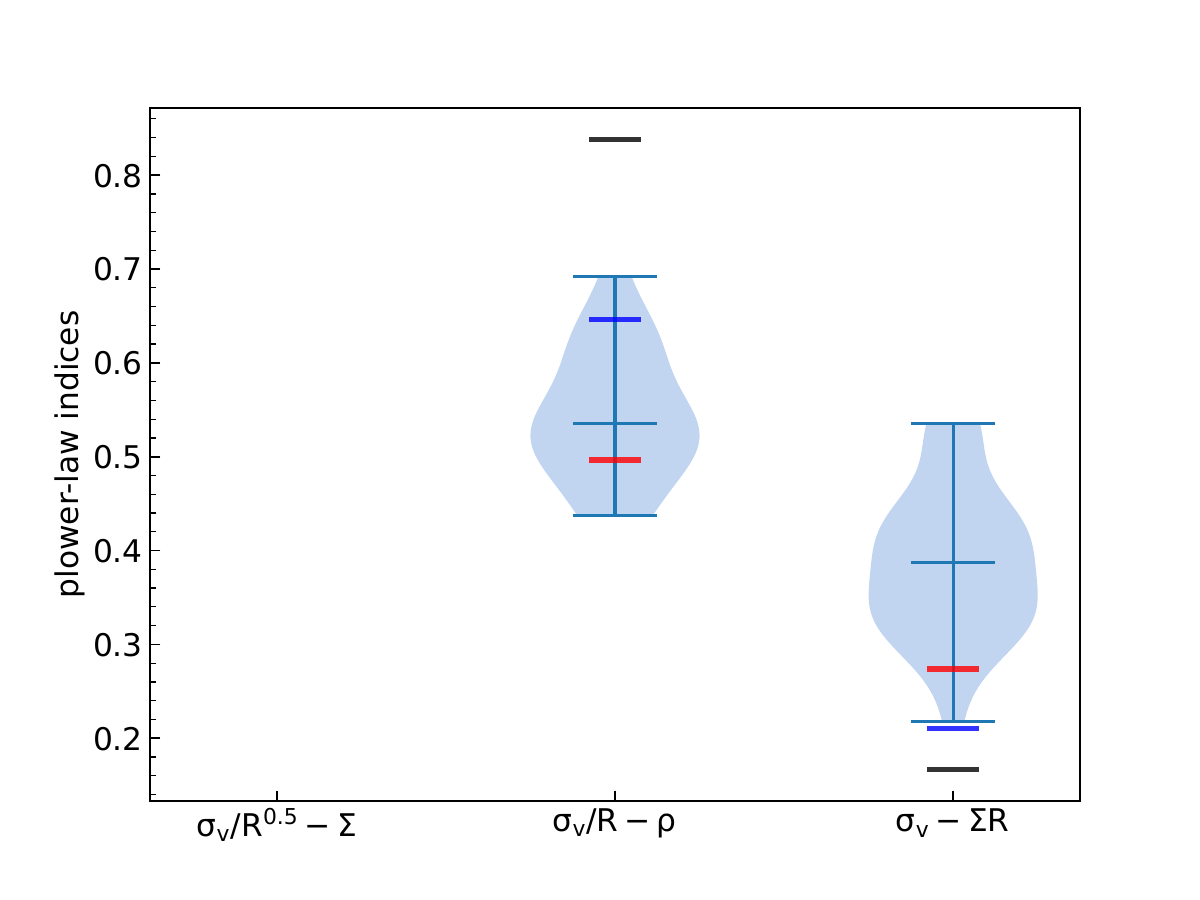}
        \\(b)
    \end{minipage}
	\caption{Same as Figure \ref{fig:pearson}, except that the relations considered here are: (1) the Keto-Heyer relation ($\sigma_v/R^{0.5}$ versus $\Sigma$), (2) the revised Keto-Heyer relation ($\sigma_v/R$ versus $\rho$), and (3) the relation between $\sigma_v$ and $\Sigma R$.}\label{fig:pearson1}
\end{figure*}

Figure \ref{fig:pearson1}(a) compares the Pearson correlation coefficients for three relations involving velocity dispersion and density, which are the relations between $\sigma_v/R^{0.5}$ and $\Sigma$ (the Keto-Heyer relation), $\sigma_v/R$ and $\rho$ (the revised Keto-Heyer relation), and the relation between $\sigma_v$ and $\Sigma R$. It can be seen that the revised Keto-Heyer relation exhibits strong correlations in both the type 2 and type 4 scenarios. Compared to the original Keto-Heyer relation, the revised version exhibits systematically higher correlation coefficients, with a median of 0.73 approaching the maximum value (0.77) observed in the original relation. Moreover, its correlation is comparable to that of the $\sigma_v$–$\Sigma R$ relation.

It is worth noting that, since all the physical parameters involved in the two correlations we compare are related to radius, the higher correlation of the revised Keto-Heyer relation may partially arise from the stronger correlated scaling between the two physical parameters and $R$. Experiments have been done to test this conjecture. Using samples randomly bootstrap-sampled from real cloud data, we found that when $R$, $M$, and $\sigma_v$ are all sampled independently, the Pearson correlation coefficient between $\log{(\sigma_v/R)}$ and $\log{\rho}$ (0.74) is indeed higher than that between $\log{(\sigma_v/R^{0.5})}$ and $\log{\Sigma}$ (0.52). However, when we take into account the correlations among the physical parameters of molecular clouds by jointly sampling $M$ and $\sigma_v$ while sampling $R$ separately, the difference between the two correlation coefficients becomes considerably smaller. These tests indicate that the intrinsic correlation arising from the shared dependence on $R$ contributes to the enhanced correlation in the revised Keto–Heyer relation, although it does not appear to fully account for the observed improvement.

Figure \ref{fig:pearson1}(b) presents the power-law indices of the type 4 relations, in which the Keto-Heyer relation is not shown due to the lack of a stable correlation. The median value of the power-law indices for the revised Keto-Heyer relation is 0.54, which closely approximates the value of 0.5 expected for the case of a constant virial parameter. The power-law indices of the revised Keto-Heyer relation exhibit a narrower distribution compared to the relation between $\sigma_v$ and $\Sigma R$, with minimum and maximum values of 0.44 and 0.69, respectively.

\begin{figure*}[!htb]
	\centering
	\begin{minipage}[t]{0.49\linewidth}
		\centering
		\includegraphics[width=\linewidth]{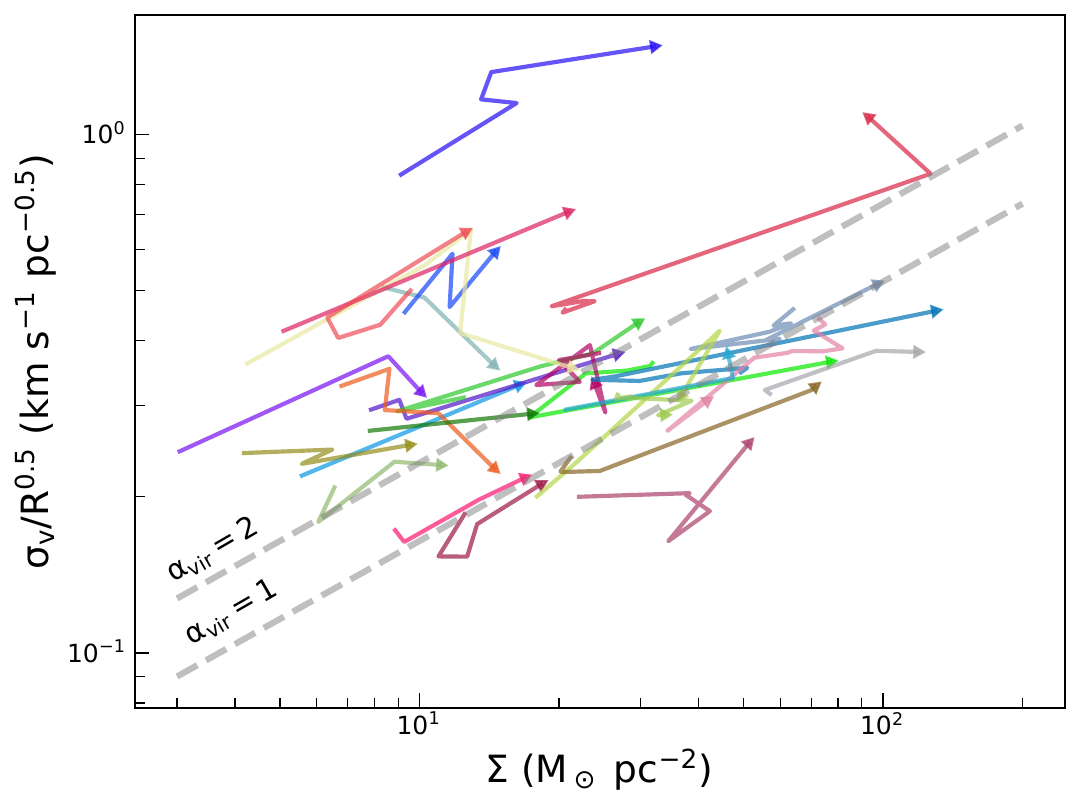}
		\\(a)
	\end{minipage}
	\begin{minipage}[t]{0.49\linewidth}
		\centering
		\includegraphics[width=\linewidth]{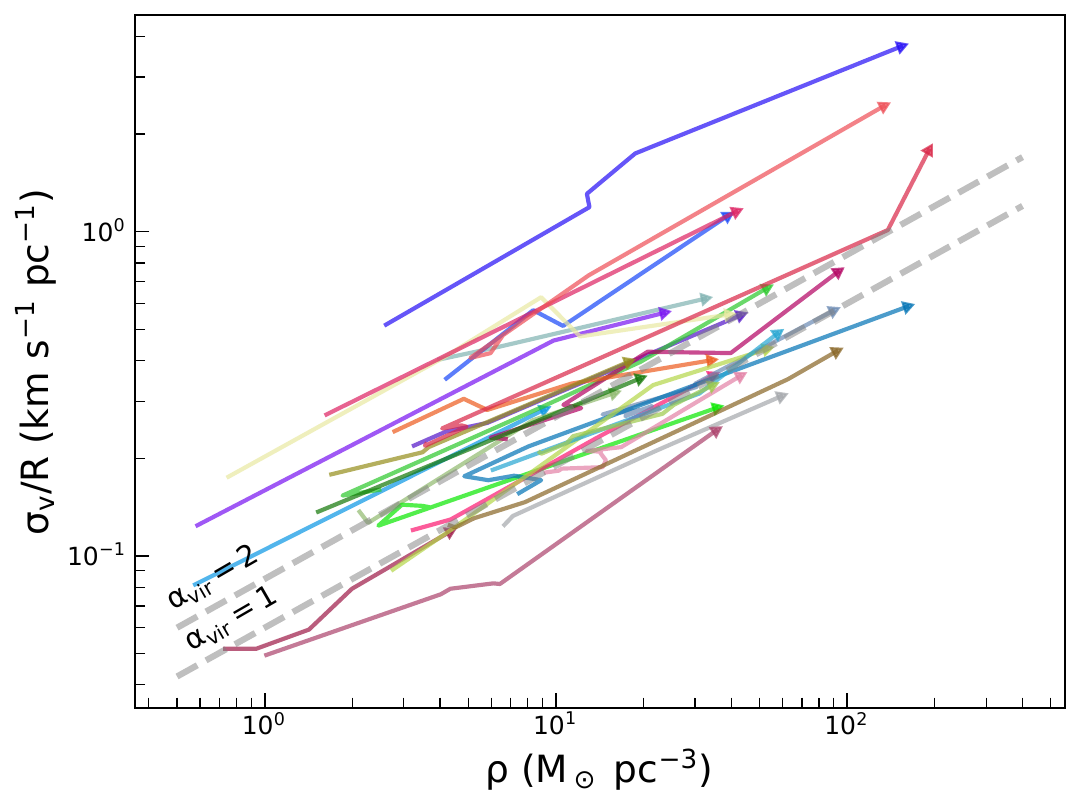}
		\\(b)
	\end{minipage}
	\caption{(a) Trajectories of molecular clouds and their substructures in the Keto-Heyer diagram. The arrow positions represent the leaf structures with the highest height in each cloud. The trajectories are plotted by iteratively tracing their parent structures, and the arrow directions indicate the progression from parent structures to descendant structures. (b) Same as (a) but in the $\sigma_v/R - \rho$ diagram.}\label{fig:trajectory}
\end{figure*}

We further illustrate this issue by examining the trajectories of the structures in the virial diagram. The differential virial analysis method proposed by \cite{Krumholz2025} suggests that the relative variation of the virial parameter between adjacent structures is a more reliable indicator of gravitational binding than its absolute value, due to uncertainties in the derivation of the virial parameter. Our methodology shares certain conceptual similarities with the differential virial analysis method proposed by \cite{Krumholz2025} and minimizes the influence of absolute error in the calculation of the virial parameter.
Figure \ref{fig:trajectory} presents the trajectories of substructures of different molecular clouds in the Keto-Heyer diagram and the $\sigma_v/R-\rho$ diagram. We selected molecular clouds comprising more than 15 substructures, where the Pearson correlation coefficient between $\log{(\sigma_v/R)}$ and $\log{\rho}$ for their substructures exceeds 0.7. Each colored arrow corresponds to the leaf structure with the highest height (i.e., the most compact structure) in a given cloud. The trajectories are constructed by iteratively tracing their parent structures. It can be seen that many clouds which exhibit poor correlation in the Keto-Heyer relation display a strong correlation in the $\sigma_v/R-\rho$ relation. Furthermore, in panel (a), the trajectories often show a turnover, while the trajectories in panel (b) exhibit significantly better monotonicity. We can see intuitively that in panel (b), the trajectories tend to be distributed along the line of constant virial parameter, i.e., the power-law indices of the type 4 revised Keto-Heyer relations are around 0.5
. Therefore, we suggest that the low correlation coefficients of the type 4 Keto-Heyer relation in Figure \ref{fig:pearson}(a) do not necessarily indicate that the substructures of molecular clouds are inconsistent with the predictions of the GHC model. Combined with the trajectories in Figure \ref{fig:trajectory}(b), which run parallel to the line of constant virial parameter, we suggest that the hierarchical structures within molecular clouds follow a revised form of the Keto-Heyer relation. Although these trajectories exhibit vertical shifts compared to the line corresponding to a virial parameter of 2, this may be due to distance measurement uncertainties or discrepancies between the assumptions in the virial parameter calculation and actual conditions. Overall, these substructures are consistent with predictions from the GHC model.

\subsection{Comparison with MHD Simulation} \label{sec:4.3}
In addition to scaling relations, models dominated by gravity and turbulence also differ in the formation mechanisms of local dense structures. In the TS model, the substructures within molecular clouds are formed by the compression of supersonic shocks generated by turbulence and, as a result, their density, $\rho$, should approximately follow the Rankine-Hugoniot relation for strong isothermal shocks: $\rho \approx M_\mathrm{s}^2\rho_{bg}$. Therefore, molecular clouds having higher $M_\mathrm{s}$ are expected to exhibit more hierarchical internal structures with higher densities.
This conclusion is also supported by the magnetohydrodynamic (MHD) turbulence simulation, as can be seen from Figure 4 of \cite{Burkhart2013}, in which they conducted various MHD simulations with different sets of sonic Mach numbers and Alfvénic Mach numbers, along with synthetic observations, to generate a series of simulated PPV cubes. By analyzing these cubes with Dendrogram, they found that the number of structures identified by Dendrogram, $N_\mathrm{stru}$, exhibits a power-law relation related to the $\mathrm{min\_delta}$ parameter. The $\mathrm{min\_delta}$ parameter of the Dendrogram algorithm sets the minimum brightness temperature contrast required for a leaf structure, which effectively traces the density contrast between local dense structures and the background, i.e., $\rho /\rho_{bg}$. 
Therefore, in numerical simulations with identical initial density conditions, an increase in the sonic Mach number enhances turbulent compressibility. This results in a broader range of density contrasts and a greater number of hierarchical substructures in the resulting PPV cubes. When identified using dendrogram analysis, this naturally yields a larger number of substructures. As shown by \cite{Burkhart2013}, the number of identified structures follows a power-law dependence on $\mathrm{min\_delta}$; higher Mach numbers increase the number of structures detected at a given $\mathrm{min\_delta}$, resulting in a systematic flattening of the $N_{\rm stru}$-$\mathrm{min\_delta}$ relation.

Similar experiments were performed on observational data to test whether the power-law relation between $N_{\rm stru}$ and min$\_$delta, as predicted by \citet{Burkhart2013} and found in MHD simulations without self-gravity, also holds for real molecular clouds. Although observational data do not provide access to initial conditions, different molecular clouds can be treated as independent realizations with varying physical states.

\begin{figure*}[!htb]
	\centering
	\begin{minipage}[t]{0.7\linewidth}
		\centering
		\includegraphics[width=\linewidth]{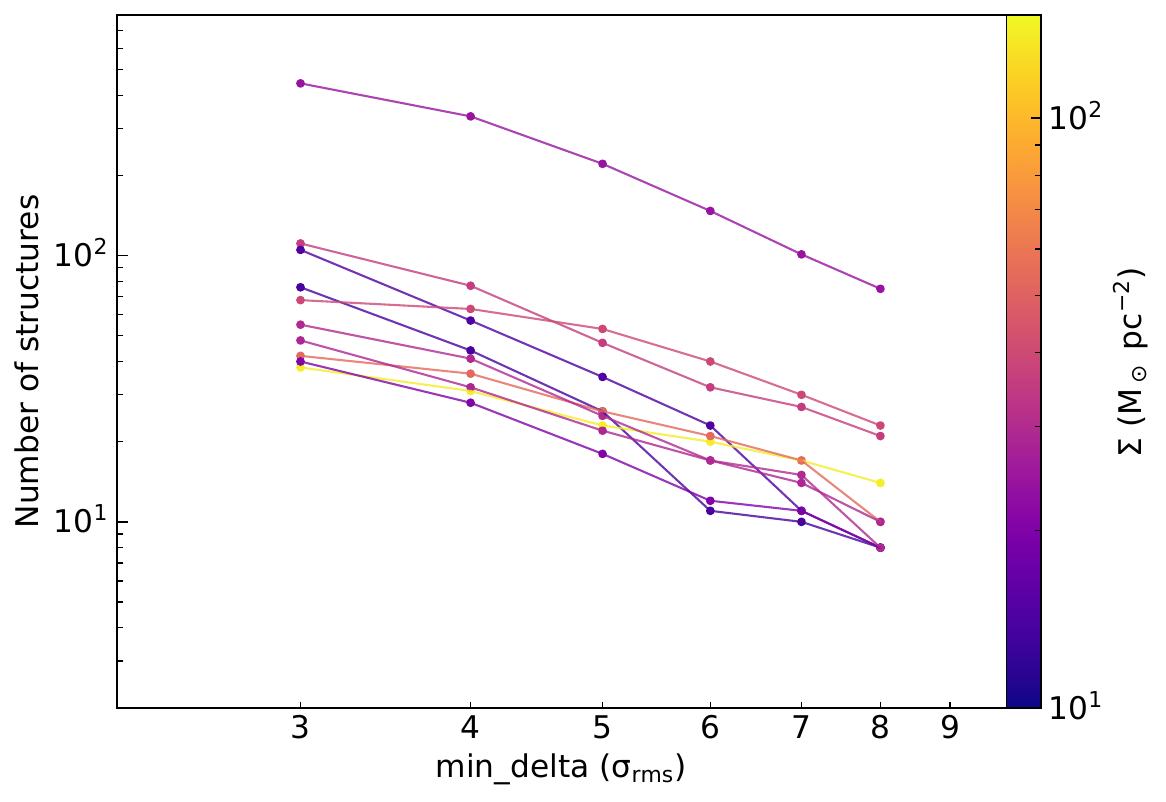}
		\\(a)
	\end{minipage}
	\begin{minipage}[t]{0.49\linewidth}
		\centering
		\includegraphics[width=\linewidth]{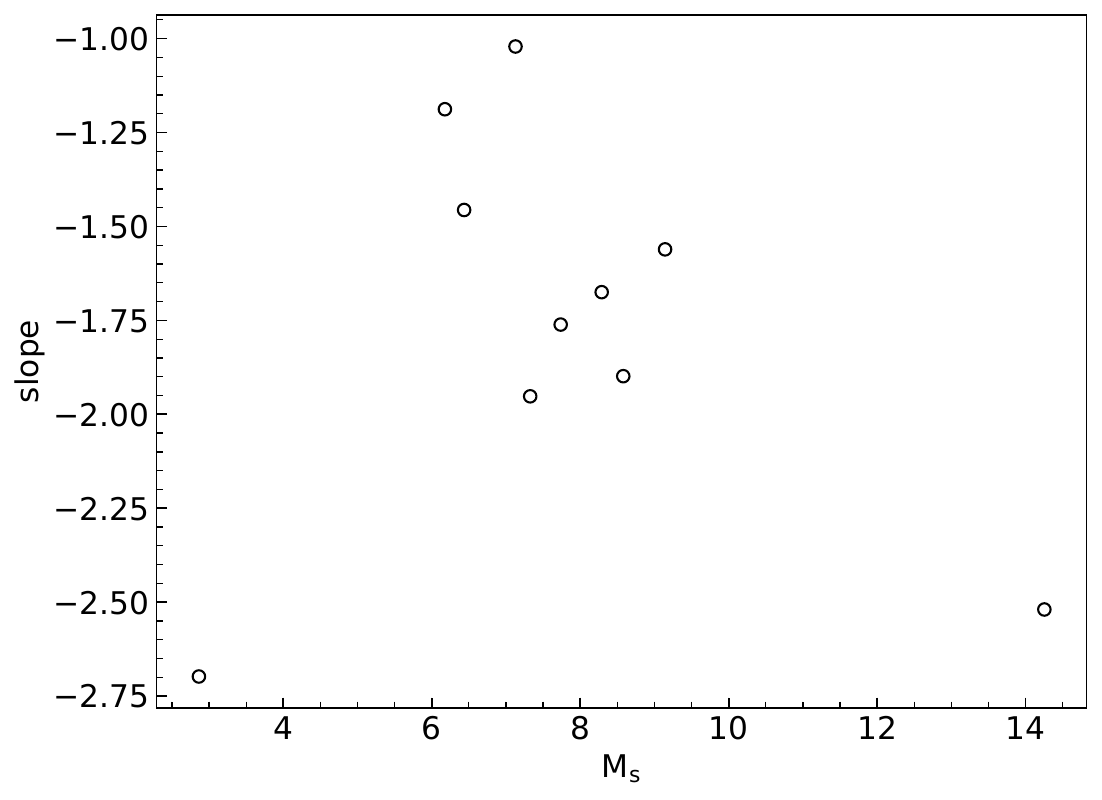}
		\\(b)
	\end{minipage}
	\begin{minipage}[t]{0.49\linewidth}
		\centering
		\includegraphics[width=\linewidth]{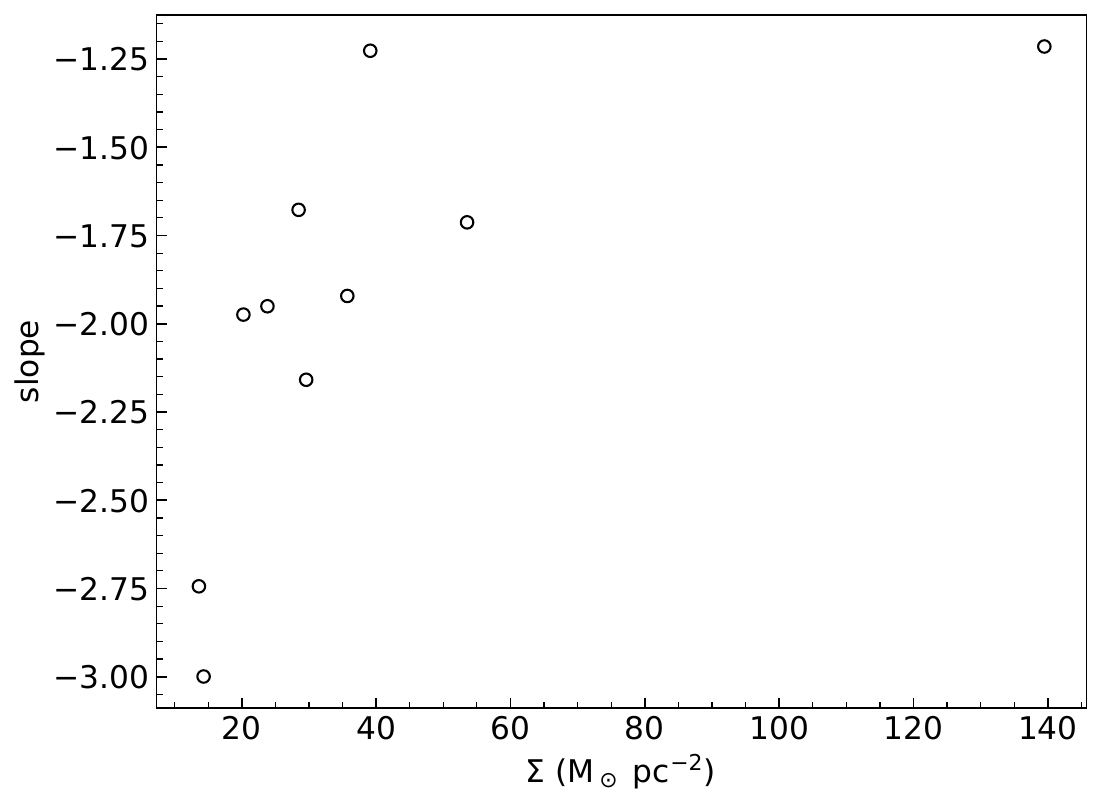}
		\\(c)
	\end{minipage}
	\caption{(a) The relation between the number of strutures identified by Dendrogram versus the selection of min$\_$ delta parameter. (b) The relation between the power-law exponent of each line in panel (a) and the sonic Mach number of the cloud. (c) Same as (b) but the x-axis stands for the surface density of the molecular cloud.}\label{fig:N_delta}
\end{figure*}

We therefore selected sufficiently complex molecular clouds in our region—specifically, those containing at least 40 substructures identified with a $\mathrm{min\_delta}$ threshold of 3$\sigma_\mathrm{RMS}$—to ensure a meaningful characterization of the $N\_\mathrm{stru}-\mathrm{min\_delta}$ relation. Figure \ref{fig:N_delta}(a) shows the resulting relation for each cloud. A power-law is fitted individually, and Figures \ref{fig:N_delta}(b) and (c) present the dependence of the fitted power-law index on the sonic Mach number and the surface density of the molecular cloud, respectively. In contrast to the predictions from the TS model and MHD simulations without self-gravity, the power-law index does not decrease with increasing sonic Mach number. Instead, it shows a strong correlation with the surface density of the molecular cloud. Because surface density directly regulates the strength of self-gravity, this behavior is more consistent with expectations from a gravitationally dominated scenario.


\section{Summary and Conclusions}\label{sec:5}
In this work, we utilized the high-sensitivity $^{12}$CO and $^{13}$CO $J=1-0$ data from the MWISP survey to investigate the hierarchical structure and dynamical state of molecular clouds in the second quadrant of the Galactic midplane ($139\fdg75 \le l \le 159\fdg75$, $-5\fdg25 \le b \le 8\fdg25$). Using a cloud-identification procedure combining DBSCAN and a non-binary Dendrogram algorithm, together with constraints on single-velocity components in $^{13}$CO integrated spectra, we identified a total of 2,912 $^{13}$CO clouds, among which 431 contain hierarchical substructures.

Based on this sample, we carried out both type 2 (multi-cloud) and type 4 (single-cloud) scaling analyses. Our main conclusions are as follows:

\begin{enumerate}

\item 
The type 2 scaling relations exhibit statistical differences among spiral arms. In the Local Arm, the linewidth–size relation follows $\sigma_v \propto R^{0.32}$, whereas for molecular clouds in the Perseus and Outer Arms, the scaling index is close to $\sim0.42$. In the Keto–Heyer diagram, molecular clouds in the Local Arm show little correlation between $\sigma_v/R^{0.5}$ and $\Sigma$, and tend to lie away from the lines of constant virial parameter, suggesting that additional external pressure may be required to maintain equilibrium. In contrast, molecular clouds in the Perseus and Outer Arms exhibit a positive Keto–Heyer correlation with a power-law index of $\sim0.3$. However, it should be noted that the observed differences among clouds in different spiral arms should be understood in the context of distance-dependent observational selection effects. The virial parameter is found to be anti-correlated with cloud radius, mass, and surface density. Overall, these type 2 results are consistent with previous large-scale CO surveys. 

\item 
Compared with type 2 relations, the type 4 scaling relations within individual molecular clouds generally exhibit stronger correlations, particularly for the $\sigma_v$–$R$ relation. The power-law indices derived under the type 4 scheme span a finite distribution range from cloud to cloud, and the indices obtained from type 2 analysis lie within this distribution. This suggests that cloud-scale (type 2) scaling relations represent ensemble averages over diverse internal dynamical states, while type 4 relations more directly probe the intrinsic kinematics of individual molecular clouds.

\item 
For the type 4 linewidth–size relation, the power-law index $\beta_1$ spans a wide range from $\sim0.3$ to $\sim0.8$, rather than stabilizing at the canonical value of 0.5. Moreover, $\beta_1$ exhibits a positive correlation with the sonic Mach number of the host cloud, such that more supersonic molecular clouds tend to display steeper linewidth–size relations. This result indicates that the linewidth–size relation is not a universal outcome of a scale-free Burgers-like turbulent cascade, but is instead modulated by the compressibility of the cloud, implying a nontrivial coupling between turbulence and gravity.

\item 
The hierarchical substructures identified within individual molecular clouds do not follow the conventional surface-density-based Keto–Heyer relation. However, when reformulated in terms of volume density (i.e., $\sigma_v/R$ versus $\rho$), the correlation improves substantially. Although intrinsic correlation arising from the shared dependence on $R$ may contribute to the enhanced correlation, it does not appear to fully account for the observed improvement. In this revised diagram, the evolutionary trajectories of substructures tend to align approximately with lines of constant virial parameter. Although some trajectories show vertical shifts relative to the line corresponding to a virial parameter of 2, which may arise from uncertainties in distance measurements or discrepancies between the assumptions and actual conditions, our results overall remain consistent with predictions from the GHC model.

\item 
Following the framework of \citet{Burkhart2013}, we examined the relation between the number of substructures and the $\rm min\_delta$ parameter. Although a power-law dependence is observed, its slope does not increase with sonic Mach number, in contrast to predictions from turbulence-dominated MHD simulations without self-gravity. Instead, the slope correlates strongly with the mean surface density of the substructures. This behavior suggests that the complexity of hierarchical structure is more closely linked to gravitational binding than to turbulent compressibility, emphasizing the importance of self-gravity in the formation of hierarchical structures within molecular clouds.

\end{enumerate}

\begin{acknowledgments}
We thank the anonymous referee for the constructive comments that helped to improve this manuscript. S.H. and Y.M. thank Professor V\'{a}zquez-Semadeni for many of his precious insights. This research made use of the data from the Milky Way Imaging Scroll Painting (MWISP) project, which is a multiline survey in $^{12}$CO/$^{13}$CO/C$^{18}$O along the northern Galactic plane with the PMO-13.7 m telescope. We are grateful to all the members of the MWISP working group, particularly the staff members at the PMO-13.7 m telescope, for their long-term support. MWISP was sponsored by National Key R$\&$D Program of China with grant 2023YFA1608001, 2017YFA0402701 and by CAS Key Research Program of Frontier Sciences with grant QYZDJ-SSW-SLH047. X.C. acknowledges the support of NSFC grant 12041305 and the support from the Tianchi Talent Program of Xinjiang Uygur Autonomous Region. Y.M. acknowledges the support of NSFC grant 12303033. H.W., Y.M., and M.Z. acknowledge the support of NSFC grant 12473026.  This research made use of astrodendro, a Python package to compute dendrograms of astronomical data (http://www.dendrograms.org/). 
\end{acknowledgments}

\clearpage
\appendix


\section{The Influence of Velocity Gradient}\label{appA1}
\begin{figure*}[!ht]
	\centering
	\begin{minipage}[t]{0.3\linewidth}
		\centering
		\includegraphics[width=\linewidth]{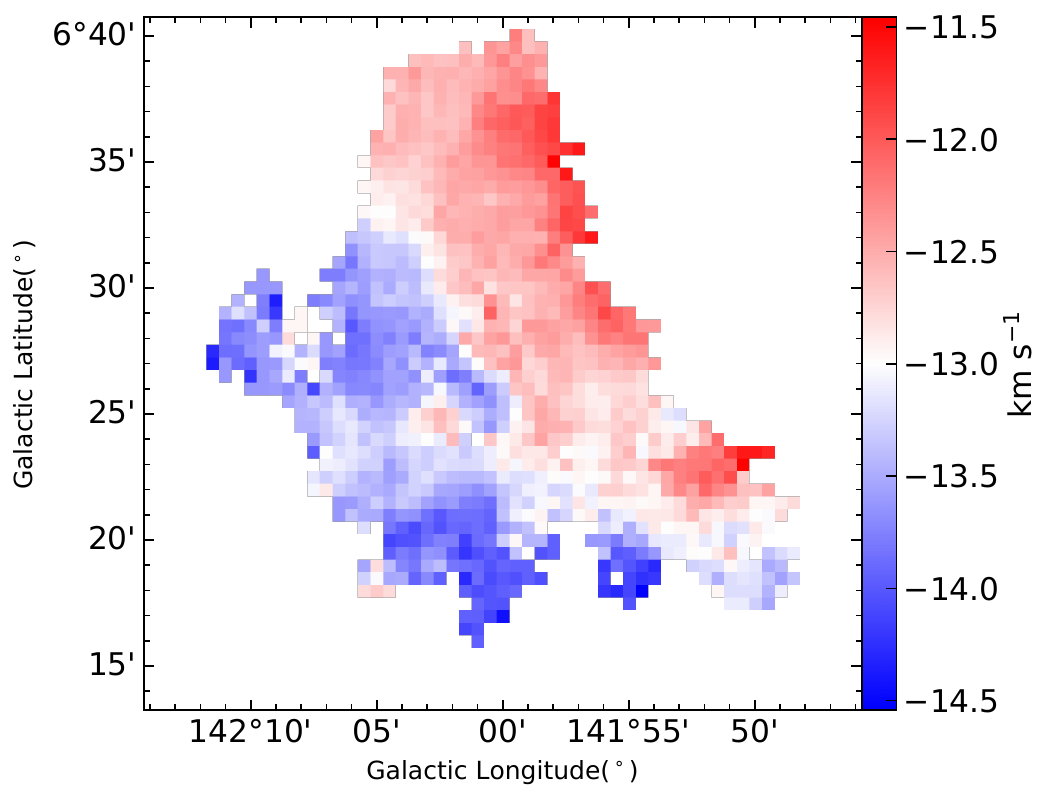}
		\\(a)
	\end{minipage}
	\begin{minipage}[t]{0.2835\linewidth}
		\centering
		\includegraphics[width=\linewidth]{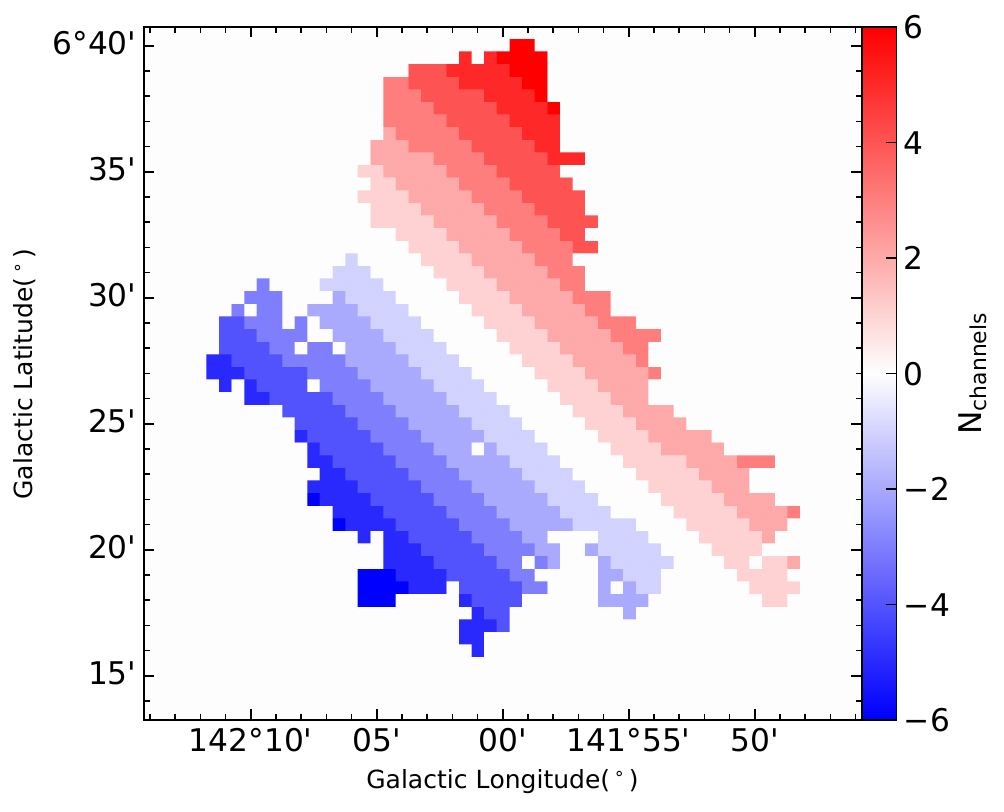}
		\\(b)
	\end{minipage}
	\begin{minipage}[t]{0.3\linewidth}
		\centering
		\includegraphics[width=\linewidth]{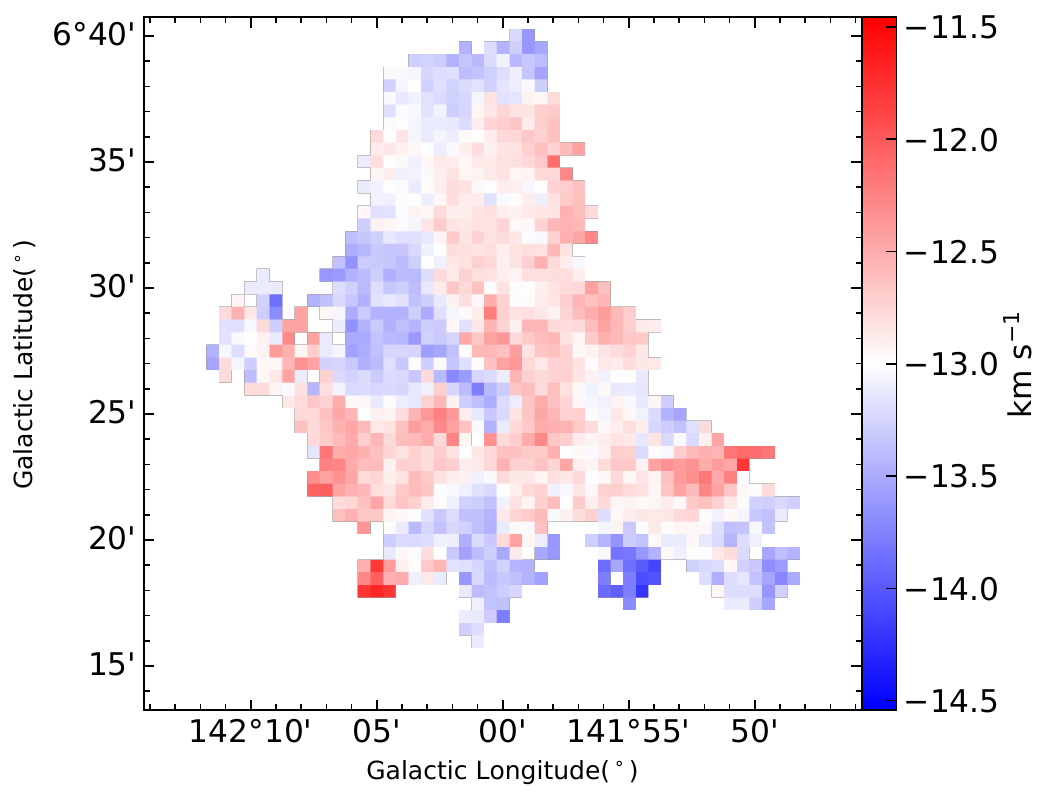}
		\\(c)
	\end{minipage}
	\caption{Schematic diagram illustrating the removal of the two-dimensional velocity gradient. (a) Centroid velocity map of the molecular cloud; (b) Channel number map corresponding to the fitted two-dimensional velocity gradient; (c) Centroid velocity map of the molecular cloud after shifting the spectrum along each line of sight by the corresponding number of channels in the reverse direction.}\label{fig:remove_gradient}
\end{figure*}

Molecular clouds may exhibit large-scale velocity gradients, which could introduce errors in the estimation of the velocity dispersion in both the molecular clouds and their substructures. In this section, we discuss the impact of removing these velocity gradients on the results. The procedure for removing velocity gradients is as follows: (1) Generate a centroid velocity map from the $^{13}$CO data of each cloud. (2) Fit the centroid velocity map to derive a two-dimensional velocity gradient. (3) Convert the velocity gradient map into a channel number map, using the $^{13}$CO channel width of $\rm \sim 0.17\ km\ s^{-1}$. (4) Shift the spectrum at each line of sight by the corresponding number of channels. Figure \ref{fig:remove_gradient} shows an example of removing the velocity gradient from a molecular cloud. Panels (a), (b), and (c) present the centroid velocity map before removing the velocity gradient, the velocity channel map corresponding to the fitted velocity gradient, and the centroid velocity map after removing the velocity gradient, respectively.

\begin{figure*}[!h]
	\centering
	\begin{minipage}[t]{0.38\linewidth}
		\centering
		\includegraphics[width=\linewidth]{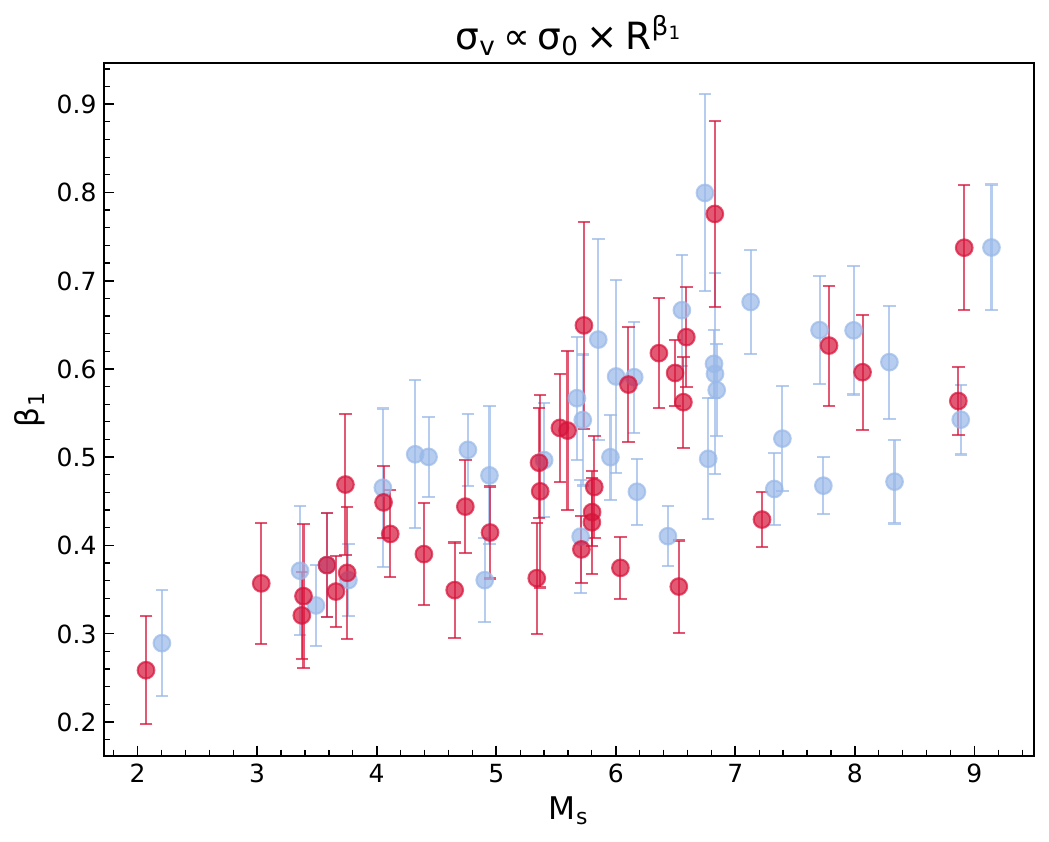}
		\\(a)
	\end{minipage}
	\begin{minipage}[t]{0.38\linewidth}
		\centering
		\includegraphics[width=\linewidth]{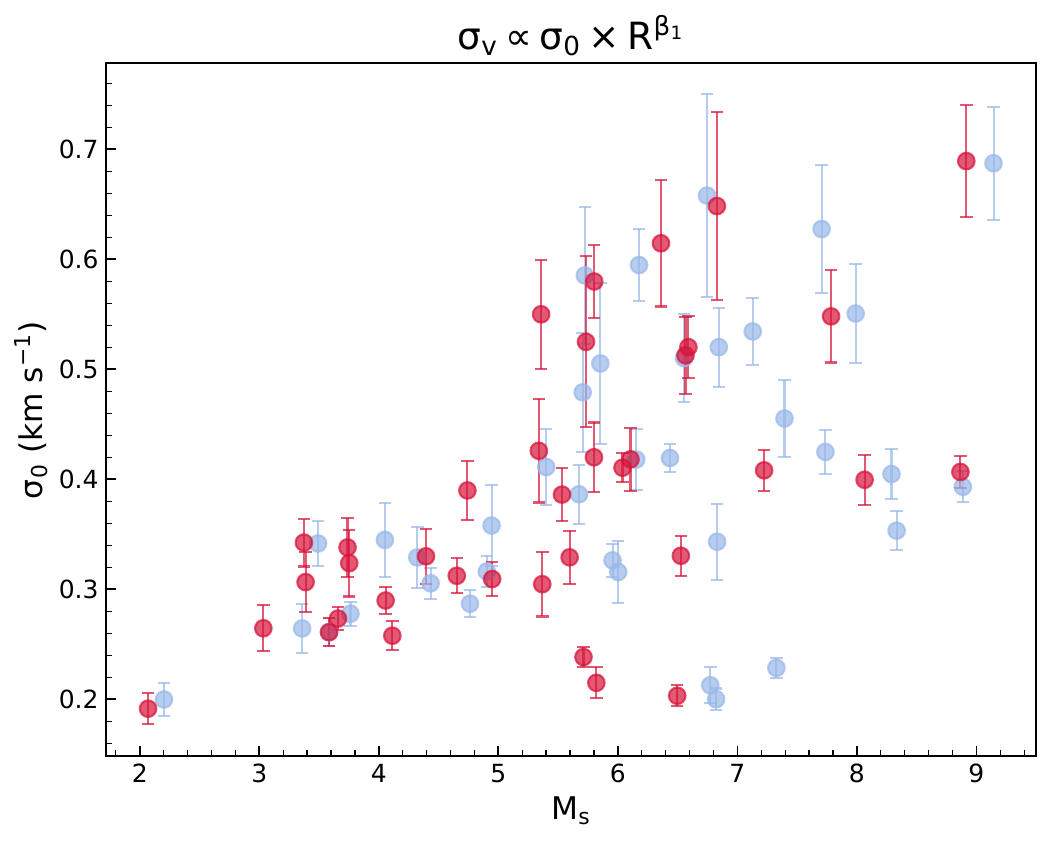}
		\\(b)
	\end{minipage}
	\caption{Same as Figure \ref{fig:beta_Ms}(a) and (b), but blue represents the fitting without removing the velocity gradient, and red represents the fitting results after removing the velocity gradient.}\label{fig:larson_vg}
\end{figure*}

Figure \ref{fig:larson_vg} shows the variation of $\sigma_0$ and $\beta_1$ of the type 4 linewidth-size relation $\sigma_v\propto \sigma_0 \times R^{\beta_1}$ as a function of the sonic Mach number of the molecular cloud. The blue points represent the fitting results without removing the velocity gradient, while the red points represent the results after the two-dimensional velocity gradient has been removed. The velocity dispersion of the molecular clouds decreases following the removal of the velocity gradient. However, the positive correlation between $\beta_1$ and $M_\mathrm{s}$ remains robust.

\clearpage

\section{Effect of Voxel Reuse on the Type 4 Linewidth-Size Relation}\label{appB}
In Section \ref{subsec:4.1}, we found that the type 4 linewidth--size relation shows a higher degree of correlation than the type 2 relation. A possible factor affecting the measured correlation is the statistical dependence among data points introduced by the hierarchical nesting of Dendrogram structures. Specifically, the samples of the type 4 linewidth--size relation are derived from hierarchical substructures within a molecular cloud. For structures located at the upper levels of the dendrogram, some of their voxels are also included in their parent structures, meaning that the corresponding data points are not fully independent of one another. In this section, we examine whether the higher correlation in the type 4 linewidth--size relation persists when voxel reuse is minimized.

\begin{figure*}[!htb]
	\centering	
    \includegraphics[width=0.9\textwidth]{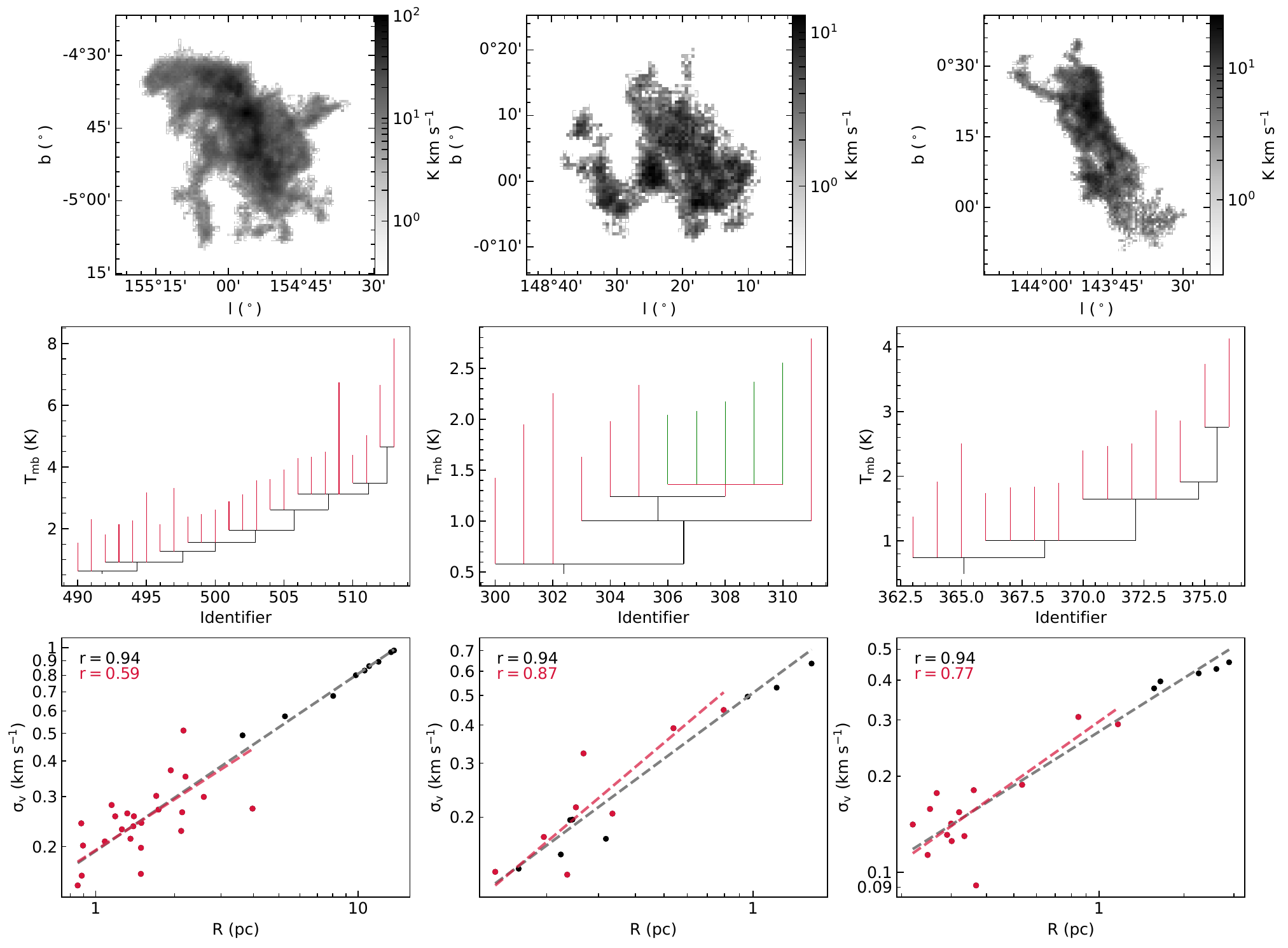}
	\caption{Schematic diagram illustrating the selection of a maximal non-overlapping substructure subset across three molecular clouds. The selected samples are identical to those in Figure \ref{fig:cases}, with the difference that in the middle row, the red structures denote those included in the subset, whereas the green structures are substructures of the red ones. Although their voxels are used as part of their parent structure, these green structures themselves are excluded from the fitting of the type 4 linewidth–size relation. In the bottom row, the red points and the dashed line indicate the selected subset and the corresponding fitting result, respectively. The Pearson correlation coefficients for all substructures and for this subset are provided in the upper‑left corner.}\label{fig:choptrunk}
\end{figure*}

We start by defining a maximal non-overlapping subset of substructures within a molecular cloud---that is, a set of substructures in which no voxel is reused and no additional substructure can be included without introducing voxel overlap. In practice, we compile all substructures within a molecular cloud into a list, from which we randomly draw one substructure and add it to the desired subset. We then remove from the list not only the drawn substructure but also all of its ancestors and descendants---that is, all structures that share any voxel with it. This process is repeated until the list is emptied. To avoid selecting subsets with an excessively reduced dynamical range in physical scale, we additionally required that the logarithmic dynamic range in radius of the selected subset be at least 0.7 times that of the full hierarchical sample. A schematic diagram illustrating the extraction of a maximal non-overlapping subset is shown in Figure \ref{fig:choptrunk}.

The extraction of a non-overlapping subset generally results in a decrease in the correlation coefficient. As illustrated by the leftmost example in Figure \ref{fig:choptrunk}, while the linewidth–size scaling relations derived from the subset and from the full set are essentially consistent, the correlation coefficient declines from 0.94 to 0.59, owing to the exclusion of large-scale, high-velocity dispersion structures at the lower levels of the dendrogram. For each of the 50 clouds containing more than 15 substructures, we construct a non-overlapping subset of Dendrogram structures following the procedure described above and record the corresponding correlation coefficient of the type 4 linewidth--size relation. This random selection procedure is repeated 2,000 times for each cloud.

\begin{figure*}[!htb]
	\centering	
    \includegraphics[width=0.7\textwidth]{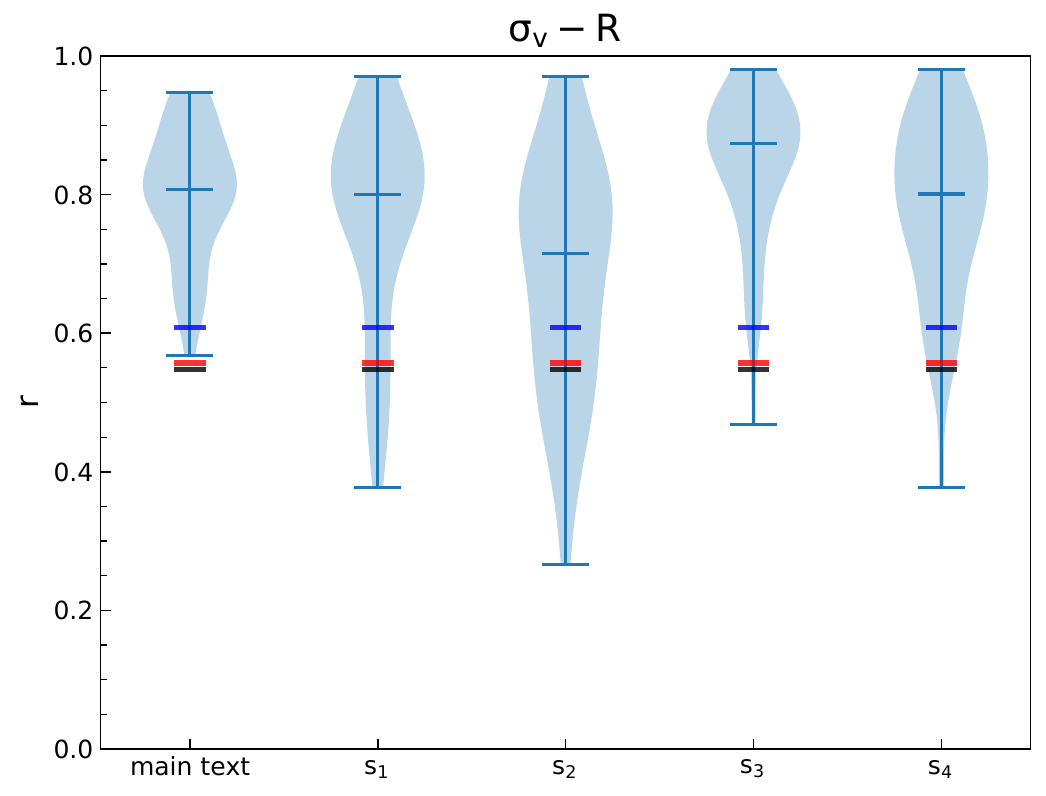}
	\caption{Violin plots of the Pearson correlation coefficients between the logarithms of $\sigma_v$ and $R$ for different samples. The leftmost column shows the results obtained using the method adopted in the main text. $\mathrm{s}_1$ and $\mathrm{s}_2$ show the maximum and median correlation coefficients, respectively, among the 2,000 non-overlapping subsets generated for each molecular cloud. $\mathrm{s}_3$ and $\mathrm{s}_4$ are the same as $\mathrm{s}_1$ and $\mathrm{s}_2$, except that the point representing the entire cloud is added to each subset to recover the full dynamic range. The short horizontal lines with colors represent the type 2 Pearson correlation coefficients among clouds in different spiral arms, consistent with Figure~\ref{fig:pearson}.}\label{fig:test_r}
\end{figure*}

Figure \ref{fig:test_r} presents the distributions of the type 4 linewidth--size correlation coefficients under several different scenarios, where $\mathrm{s}_1$ and $\mathrm{s}_2$ correspond to the maximum and median correlation coefficients, respectively, among the 2,000 randomly generated non-overlapping subsets for each cloud. As expected, removing overlapping structures moderately shifts the correlation coefficient distribution toward lower values compared with that obtained from the full samples in the main text. Nevertheless, the median value of $\mathrm{s}_2$ is 0.71, and 68\% of the values exceed 0.6, which is still higher than the type 2 correlation coefficients. 

Since the point representing the entire cloud cannot be included in any subset, the dynamic range of linewidth and size for the subset is inevitably smaller than that of the full sample. We therefore also add the point representing the entire cloud to each subset, thereby restoring the full dynamic range while introducing only limited voxel reuse. The distribution of the corresponding maximum and median correlation coefficients for the selected molecular clouds are labeled $\mathrm{s}_3$ and $\mathrm{s}_4$, respectively. In these cases, the correlations are largely recovered.

These results indicate that the hierarchical dependence of Dendrogram structures does contribute to the enhanced type 4 correlation coefficients, but it is not the primary origin of the stronger intracloud linewidth-size correlations.

\clearpage
\newpage
\bibliography{sample631}{}
\bibliographystyle{aasjournal}



\end{document}